\documentclass[10pt]{IEEEtran}

\usepackage{times}
\usepackage{theorem}
\usepackage{amsmath}
\usepackage{amssymb}
\usepackage{bm}
\usepackage{subfigure}
\usepackage{cite}
\usepackage{hyperref}
\usepackage{graphicx}
\usepackage{tikz,pgfplots}
\usepackage{microtype}
\usepackage{cleveref}
\usepackage{multirow}

\usepackage[draft]{todonotes}   
\usepackage{comment}

\usepackage{needspace}

\input{mysymbol.sty}
\usepackage{mdframed}

\newtheorem{mytheorem}{\bf Theorem}
\newtheorem{mydefinition}{\bf Definition}
\newtheorem{mylemma}{\bf Lemma}
\newtheorem{myproposition}{\bf Proposition}
\newtheorem{remark}{\bf Remark}
\newtheorem{problem}{Problem}
\newtheorem{assumption}{Assumption}

\usepackage{algorithm}
\usepackage[noend]{algpseudocode}

\title{Network Inference from Consensus Dynamics with Unknown Parameters}

\author{Yu Zhu, Michael T. Schaub, Ali Jadbabaie, Santiago Segarra \thanks{Y. Zhu and S. Segarra are with the Department of Electrical and Computer Engineering, Rice University. M. Schaub and A. Jadbabaie are with the Institute for Data, Systems, and Society, MIT. M. Schaub is also with the Department of Engineering Science, University of Oxford, UK. Emails: {\tt\footnotesize \{yz126, segarra\}@rice.edu, \{mschaub, jadbabai\}@mit.edu}. Funding: This work was supported by the European Union's Horizon 2020 research and innovation programme under the Marie Sklodowska-Curie grant agreement No 702410 (M. Schaub). A. Jadbabaie's research was supported by a Vannevar Bush Fellowship from the  Office of Secretary of Defense. A preliminary version of some of the results here appeared in~\cite{Segarra2017}.}}

\begin{document}
\maketitle
\begin{abstract}
	We explore the problem of inferring the graph Laplacian of a weighted, undirected network from snapshots of a single or multiple discrete-time consensus dynamics, subject to parameter uncertainty, taking place on the network.
	Specifically, we consider three problems in which we assume different levels of knowledge about the diffusion rates, observation times, and the input signal power of the dynamics.
	To solve these underdetermined problems, we propose a set of algorithms that leverage the spectral properties of the observed data and tools from convex optimization. 
	Furthermore, we provide theoretical performance guarantees associated with these algorithms.
	We complement our theoretical work with numerical experiments, that demonstrate how our proposed methods outperform current state-of-the-art algorithms and showcase their effectiveness in recovering both synthetic and real-world networks.
\end{abstract}

\includecomment{versiona}
\excludecomment{versionb}

\section{Introduction}\label{S:intro}

	Networks have become a fundamental tool to model systems across Science and Engineering, with applications ranging from physical to socio-economic and biological domains~\cite{Albert2002,Jackson2010,Strogatz2001}.  
	In certain cases, we may have relational data that quantifies the couplings between the system entities directly.
	For example, in transportation networks, we can measure traffic flows between different points in space. 
	However, in many instances, the true couplings between the system entities are unknown and have to be inferred from data collected from the system entities. 
	This task, which we refer to as network inference, is thus a fundamental step prior to any further network analysis.    

	Network inference has been studied from several perspectives in the literature, and different models have been adopted that associate the network topology with the observed data. 
	Historically, there are two main lines of research, which are based on statistical models and physically-motivated models~\cite{mateos_review, dong_review}, respectively.   
	A well-known statistical model is the graphical model~\cite{graphical_model}, where the network (graph) encodes conditional independence relations among random variables defined on the system entities (nodes).
	By employing the graphical model, the problem of inferring the network is thus converted to a particular estimation of the joint probability distribution of these random variables~\cite{mateos_review, dong_review}. 
	Associated algorithms include the graphical LASSO~\cite{GLasso0,GLasso1,GLasso2}, which incorporates a graph sparsity prior into the maximum likelihood estimator to recover the precision matrix of these random variables. 
	Physically-motivated models assume that the observed data is generated by some physical process on the network such as diffusion~\cite{Rodriguez2012,Rodriguez2014,Timme2014, Shahrampour2015}, and the network recovered is expected to explain the generative process of the observations. 
	
	Graph signal processing (GSP)~\cite{sandryhaila_2013_discrete,shuman_2013_emerging,ortega_2018_graph}, a fast growing research area that seeks to extend concepts and methods in classical digital signal processing to graphs, offers a new perspective to the problem of network inference.
	Works leveraging GSP tools include: 
	(i) Models based on signal smoothness~\cite{smooth0,smooth1}, where methods to infer the network topology from smooth signals are proposed via minimizing a regularized graph Laplacian quadratic form;
	(ii) Models based on causal dependency~\cite{causal0}, where algorithms are put forth to recover the network structure capturing the dependencies among time series; and 
	(iii) Models based on network diffusion~\cite{Segarra2016,Pasdeloup2018,Hilmi2018, heat_diffusion}, which solve the problem of network inference from a stationary graph process by leveraging the spectral information contained in the observations.
	Extensions to non-stationary diffusion processes have also been recently proposed~\cite{shafipour_2017_network, shafipour_2018_identifying}. For a thorough review on the topic of network inference, see~\cite{mateos_review, dong_review}.

	In this paper, we consider the problem of \emph{network inference from snapshot observations of consensus dynamics}. 
	Consensus has been one of the most popular and well-studied dynamics on networks~\cite{consensus1,consensus2,consensus3} due to both its analytic tractability and its simplicity in approximating several fundamental behaviors.
	For example, in socio-economic domains, consensus provides a model for opinion formation in social networks.
	For engineering systems, it has been considered as a basic building block for an efficient distributed computation of global functions in networks of sensors, robots, or other agents.

	We consider snapshot data since this is a common scenario in modern observational datasets. 
	For instance, in the study of gene-expression via single-cell RNA sequencing~\cite{Saliba2014}, we can only obtain snapshot data, as the process of obtaining a sample destroys the cell under consideration. 
	To obtain multiple samples, we thus have to replicate the experiments with similarly prepared cells. 
	Another example is the monitoring of ecological populations from abundance data~\cite{Gray2014}.
	While the underlying process may well be continuous in time, very often we can only access snapshot information at specific instances of time.
	
\vspace{1mm}
\noindent \emph{Contributions:} 
	We study the problem of inferring a network topology from snapshots of discrete-time consensus dynamics with parametric uncertainty.
	Our specific contributions can be summarized as follows: (i) We formulate three problems (Problems~\ref{P:single}-\ref{P:multiple} in Section~\ref{S:pre_and_problem}) with increasing degree of uncertainty about the parameters of the dynamics, and propose algorithms to solve them; (ii) We provide provable bounds on the performance of key steps within these algorithms; and (iii) We illustrate the performance of the proposed methods and compare it with state-of-the-art solutions for synthetic and real-world settings.

The proposed observation model strikes a balance between being specific and versatile.
On the one hand, the advocated model is more specific than those assuming signal smoothness \cite{smooth0,smooth1} or generic diffusion processes~\cite{Segarra2016,Pasdeloup2018}. 
This enables us to obtain better network recovery performance when the model (approximately) holds.
On the other hand, our formulation with unknown filter parameters (i.e., Problem~\ref{P:multiple}) includes a wider range of settings compared to those assuming a single underlying stationary graph process~\cite{Segarra2016,Pasdeloup2018,Hilmi2018} or specific functional forms for the dynamics~\cite{heat_diffusion}. 
For our Problems~\ref{P:single} and \ref{P:constant_unknown} which present scenarios with less parameter uncertainty compared to the more general Problem \ref{P:multiple}, the work \cite{Hilmi2018} is closest to our work.
However, the proposed solutions differ significantly to those in \cite{Hilmi2018} as discussed throughout the paper and illustrated numerically in Section~\ref{S:num_exp}.	

\vspace{1mm}\noindent	\emph{Paper outline:}
	The remainder of this article is structured as follows.
	Preliminary concepts related to GSP and consensus are reviewed in Section~\ref{S:pre_and_problem}. 
	In Section~\ref{S:problem_set}, we give a formal account of our problem setup and introduce three concrete problem formulations.
	In Section~\ref{S:single} we provide a detailed analysis of the first problem along with the algorithm that we propose for its solution.
	Section~\ref{S:multiple} builds upon insights gained by studying the first problem, and provides solutions to the other two problems considered.
	Numerical experiments based on both synthetic and real-world data are presented in Section~\ref{S:num_exp}, and closing remarks are included in Section~\ref{S:conclusions}.
	
	\vspace{1mm}\noindent
	\emph{Notation:} 
	The entries of a matrix $\mathbf{X}$ and a vector $\mathbf{x}$ are denoted by $X_{ij}$ and $x_i$, respectively;
	to avoid confusion, the alternative notation $[\mathbf{X}]_{ij}$ and $[\mathbf{x}]_{i}$ will be used occasionally, when dealing with indexed families of matrices and vectors.
	Operations $(\cdot)^{\top}$, $(\cdot)^{\dag}$, $\mathbb{E}(\cdot)$ and $\mathbb{P}(\cdot)$ represent transpose, pseudo-inverse, expected value and probability, respectively. 
	$\mathbf{0}$, $\mathbf{1}$ and $\mathbf{I}$ refer to the all-zero vector, the all-one vector, and the identity matrix, where the sizes are clear from context.
	$\diag(\mathbf{x})$ denotes a diagonal matrix whose $i$th diagonal entry is $x_i$.
	{$\mathrm{vec}(\mathbf{X})$ stacks the columns of $\mathbf{X}$ into a single column vector.
	For any set $\mathcal{S}$, $\mathbf{x}_\mathcal{S}$ denotes the vector formed by the entries of $\mathbf{x}$ indexed by $\mathcal{S}$.
	}

\section{Preliminaries}\label{S:pre_and_problem}

We briefly introduce basic GSP concepts (Section~\ref{SS:gsp_basic}) as well as the mathematical formulation of discrete-time consensus (Section~\ref{SS:consensus}). 

\subsection{Fundamentals of graph signal processing}\label{SS:gsp_basic}

\noindent \textbf{Graphs and graph signals.}
Consider a weighted and undirected graph $\mathcal{G}$ with $N$ nodes, whose structure is encoded by the weighted adjacency matrix $\mathbf{A}\in\mathbb{R}^{N\times N}$.
If nodes $i$ and $j$ are connected, the edge weight $A_{ij}=A_{ji}>0$ reflects the strength of the connection.
If there is no edge between nodes $i$ and $j$, we have $A_{ij}=A_{ji}=0$.

A \emph{graph signal} defined on $\mathcal{G}$ can be represented as a vector $\mathbf{x}=[x_1,\cdots,x_N]^{\top}\in\mathbb{R}^N$, where $x_i\in\mathbb{R}$ denotes a scalar signal value associated with node $i$.

\vspace{1mm}

\noindent \textbf{Graph shift operator and graph filters.}
A \emph{graph shift operator} $\mathbf{S}\in\mathbb{R}^{N\times N}$ \cite{sandryhaila_2013_discrete}, is a matrix whose off-diagonal sparsity pattern is identical with the adjacency matrix: $S_{ij}$ can only be non-zero if $A_{ij}\neq 0$ or $i=j$.
Typical choices for $\mathbf{S}$ are the adjacency matrix \cite{sandryhaila_2013_discrete}, the graph Laplacian \cite{shuman_2013_emerging} and their respective generalizations. 

A \emph{graph filter} {is a map between graph signals} and is defined as a matrix function $h(\cdot)$ of a graph shift operator $\mathbf{S}$. 
An important class of graph filters are linear and shift-invariant (LSI) graph filters.
An LSI filter can be expressed as a polynomial of $\mathbf{S}$, i.e. $h(\mathbf{S})=\sum_{l=0}^{T}h_l\mathbf{S}^l$, where $T$ and $\{h_l\}$ denote the \emph{filter degree} and \emph{filter coefficients}, respectively~\cite{sandryhaila_2013_discrete, segarra_2017_filters}. 
For a given input signal $\mathbf{x}$, the output of the graph filter is given by $\mathbf{y}=h(\mathbf{S})\mathbf{x}$.

\vspace{1mm}

\noindent \textbf{The set of combinatorial graph Laplacians.}
In this paper, we concentrate on the \emph{combinatorial graph Laplacian} (CGL) as our graph shift operator, defined as $\mathbf{L}=\text{diag}(\mathbf{A1})-\mathbf{A}$. 
The set of all CGL matrices can be written as
\begin{equation}\label{E:CGL}
\mathcal{L}_c = \{\mathbf{L} \,|\, L_{ij}=L_{ji}\leq 0\text{ for } i\neq j,\, \mathbf{L1}=\mathbf{0}\}.
\end{equation}
Since $\mathbf{L}$ is a real and symmetric matrix, its eigendecomposition can be written as $\mathbf{L}=\mathbf{V\Lambda V}^{\top}$, where $\mathbf{V}$ is a unitary matrix whose columns are the eigenvectors of $\mathbf{L}$, and $\mathbf{\Lambda}=\text{diag}(\pmb{\lambda})$ collects the eigenvalues.
Notice that~\eqref{E:CGL} implies that $\mathbf{L}$ is diagonally dominant, which ensures that $\mathbf{L}$ is positive semi-definite.

Throughout the paper we assume the eigenvalues of $\mathbf{L}$, $0=\lambda_1<\lambda_2<\cdots<\lambda_N$, are distinct. 
This assumption is not fundamental from a technical viewpoint, but simplifies the presentation of our results.
In particular, it implies that $\mathcal{G}$ is connected.

\subsection{Discrete-time consensus dynamics}\label{SS:consensus}
We consider discrete-time linear consensus dynamics~\cite{consensus1,consensus2,consensus3}, evolving  on a graph $\mathcal{G}$ with Laplacian~$\mathbf{L}$:
\begin{equation}\label{E:consensus_matrix}
\mathbf{x}[t] = \mathbf{x}[t-1]-\alpha_t\mathbf{L}\mathbf{x}[t-1]= (\mathbf{I}-\alpha_t\mathbf{L})\mathbf{x}[t-1].
\end{equation}
Here the vector $\mathbf{x}[t]$ is a (time-varying) graph signal, whose entries $x_i[t]$ correspond to the opinion of agent $i$ in the network at time $t$. 
We assume that $0<\alpha_t< {\lambda_{N}^{-1}} $ for all $t$, such that the dynamics is stable~\cite{consensus2}.

The above equations describe a dynamics in which agent $i$ updates its opinion according to a linear combination of (i) its previous opinion and (ii) a weighted discrepancy with its neighbors at the previous time point.
The parameter $\alpha_t$ is called the \emph{diffusion rate} and describes the weight given to the discrepancy term in the update at time $t$.
Under the dynamics~\eqref{E:consensus_matrix}, the opinions of all agents coincide asymptotically, i.e., $\lim\limits_{t\to\infty}\mathbf{x}[t]=c\mathbf{1}$ where $c$ is a real constant.

\section{Problem formulation}\label{S:problem_set}

We study the problem of inferring the topology of a network from the observation of $M$ consensus dynamics at a single point in time.
In contrast to other identification tasks, we assume that only such \emph{snapshot} information is available, i.e., no trajectories of the states or detailed knowledge of the initial condition is available.

Within this setting, we study three different identification problems. 
The difference between these problems lies in the amount of knowledge that we assume about the dynamics.
In our first problem formulation (Problem~\ref{P:single}), we assume that the diffusion rates and observation times are known.
Subsequently, we relax these assumptions, and allow for unknown -- albeit constant -- diffusion rates and observation times (Problem~\ref{P:constant_unknown}). Finally, in Problem~\ref{P:multiple} we tackle the most general case where we observe multiple consensus dynamics with unknown and possibly different diffusion rates and observation times.

From the perspective of GSP, our task can also be expressed as learning the graph Laplacian $\mathbf{L}$ from the observations of a set of filtered graph signals $\mathbf{y}$.
Let $\bm{\xi}:=\mathbf{x}[0]$ denote the graph signal at time zero and $\mathbf{y}:=\mathbf{x}[T]$ the state of the dynamics at a specific observation time $T>0$.
From \eqref{E:consensus_matrix} we know that $\mathbf{y}$ and $\bm{\xi}$ are related as
\begin{equation}\label{E:consensus_filter}
\mathbf{y}=h(\mathbf{L})\bm{\xi}, \quad \text{where } h(\mathbf{L})=\prod_{t=1}^T (\mathbf{I}-\alpha_t\mathbf{L}).
\end{equation}
Thus the dynamical description given in~\eqref{E:consensus_matrix} can be equivalently phrased in terms of~\eqref{E:consensus_filter} from a GSP perspective.
In this paper we adopt this GSP perspective and assume that we observe a set of filtered graph signals $\{\mathbf{y}_k\}_{k=1}^M$ of the form~\eqref{E:consensus_filter}.
Before turning to the specific problem formulations we state our main assumptions.

\subsection{Assumptions}
To ensure that there is some non-trivial information about $\mathbf{L}$ contained in the observations $\{\mathbf{y}_k\}_{k=1}^M$, we assume that the dynamics has not reached asymptotic consensus. Also, as discussed in Section~\ref{SS:consensus}, we assume diffusion rates small enough to ensure convergence.

\begin{assumption}[Finite-time consensus dynamics]\label{A:dynamics}
	The observation time $T$ is finite, i.e. $T<\infty$, and the diffusion rates satisfy $0<\alpha_t< {\lambda_{N}^{-1}}$ for all $t$.
\end{assumption}

Second, we assume that we \emph{cannot} control or observe the initial input $\bm{\xi}$, but have some knowledge about its distribution.
Specifically, we make the following assumption about the unknown input signal throughout this paper.
\begin{assumption}[White Gaussian input]\label{A:input_noise}
	The initial condition $\bm{\xi}$ in~\eqref{E:consensus_filter} is a Gaussian random graph signal $\bm{\xi} \sim \mathrm{N}(\mathbf{0}, \sigma^2\mathbf{I})$, where $\sigma^2$ denotes the input power.
\end{assumption}

\subsection{Formal problem statements}

Consider a set of $M$ independent, identically distributed initial conditions $\{\bm{\xi}_k\}_{k=1}^M$ with \emph{unknown} input power $\sigma^2$.
In our most general formulation, each of these initial conditions evolve according to a consensus dynamics such that 
\begin{equation}\label{E:general_consensus_filter}
\mathbf{y}_k=h_k(\mathbf{L})\bm{\xi}_k, \quad \text{where } h_k(\mathbf{L})=\prod_{t=1}^{T_k} (\mathbf{I}-\alpha_t^{(k)}\mathbf{L}),
\end{equation}
for $k = 1, \cdots, M$.
In our first problem, we focus on the specific case where all the dynamics are identical, and the only unknown parameter is the input power $\sigma^2$.

\begin{problem}[unknown input noise level]\label{P:single} \normalfont
    Given the set of $M$ outputs $\{\mathbf{y}_k\}_{k=1}^M$ from \eqref{E:general_consensus_filter} for an {\emph{identical}} dynamics $h_k(\mathbf{L}) \equiv h(\mathbf{L})$, estimate the Laplacian $\bbL$ for \emph{known} observation time $T$ and diffusion rates $\{ \alpha_t\}$.
\end{problem}

The second problem formulation may be interpreted as a relaxation of Problem~\ref{P:single} where we do not know the observation time $T$ nor the diffusion rate $\alpha$, which we assume to be identical for all time steps {in the context of Problem~\ref{P:constant_unknown}}.

\begin{problem}[unknown graph filter]\label{P:constant_unknown} \normalfont
    Given the set of $M$ outputs $\{\mathbf{y}_k\}_{k=1}^M$ from \eqref{E:general_consensus_filter} for an {\emph{identical}} dynamics $h_k(\mathbf{L}) \equiv h(\mathbf{L})$, estimate the Laplacian $\bbL$ for \emph{unknown} observation time $T$ and constant diffusion rate $\alpha_t \equiv \alpha$.
\end{problem}

Finally, we consider the more challenging and general setting where there are $M$ different consensus dynamics with unknown parameters evolving on a single network with Laplacian $\mathbf{L}$.  

\begin{problem}[unknown set of filters]\label{P:multiple} \normalfont
    Given the set of $M$ outputs $\{\mathbf{y}_k\}_{k=1}^M$ from \eqref{E:general_consensus_filter} for {\emph{different}} dynamics $h_k(\mathbf{L})$, estimate the Laplacian $\bbL$ for \emph{unknown} observation times $T_k$ and diffusion rates $\{\alpha_t^{(k)}\}$.
\end{problem}

The three problems above may be interpreted in terms of a hierarchy of assumptions on the knowledge of the parameters.
While Problem~\ref{P:single} is already challenging as we only rely on snapshot data and have no knowledge of the exact initial conditions, in Problems~\ref{P:constant_unknown} and~\ref{P:multiple} we make even weaker assumptions on the diffusion parameters $T_k$ and $\{\alpha^{(k)}_t\}$.
However, in many applications we are precisely confronted with such strongly underdetermined problems: 
while we may have some (approximate) model of the functional form of the dynamics (such as a consensus or a diffusion), the specific parameters of such dynamics as well as their initial conditions are often not known. 
Hence, ideally we would like to infer the network together with the parameters of the dynamics.

To see the practical relevance of our problem setup, let us consider an illustrative example for Problem~\ref{P:multiple} in the context of social sciences.
Assume that we observe, at a specific point in time, the opinion profile of all agents in a social network represented by $\mathbf{L}$ regarding $M$ independent topics, each of which evolved according to a consensus dynamics as described in \eqref{E:general_consensus_filter}. 
The discussion about each of the $M$ topics, which we index by $k\in\{1,\cdots,M\}$, may have started at a different point in time -- corresponding to unknown durations $T_k$. 
Moreover, the interactions between the agents may have been heterogeneous across topics and time -- associated with unknown diffusion rates $\alpha^{(k)}_t$.
Our goal is to identify the underlying social network topology $\mathbf{L}$ from the observation of $M$ opinion profiles $\{\mathbf{y}_k\}$ at a given time point. For further illustration of these problems, including real-world data implementations, {see Section~\ref{Ss:num_exp_real}}.

Note that in Problems~\ref{P:single} and~\ref{P:constant_unknown}, $\{\mathbf{y}_k\}_{k=1}^M$ are independent realizations of the output of a \emph{single}, {possibly unknown (Problem~\ref{P:constant_unknown})} graph filter $h(\mathbf{L})$ under a white noise input. 
This is precisely the definition of a stationary graph process~\cite{stationary, perraudinstationary2016, girault2015stationary}. 
Hence, Problems~\ref{P:single} and~\ref{P:constant_unknown} can also be expressed as inferring $\mathbf{L}$ from independent samples of {an \emph{identical}} stationary graph process~$\mathbf{y}$.
This has important implications for the solution of these problems that we can exploit.
In contrast, in Problem~\ref{P:multiple} each observation corresponds to a realization of a \emph{different} stationary process, and thus requires an adjusted solution strategy.

\begin{remark}[Related work \cite{heat_diffusion}] \normalfont
    In \cite{heat_diffusion}, the authors consider a set of heat diffusion processes which can be modeled by graph filters in the form of $h(\mathbf{L})=\exp(-\tau\mathbf{L})$ with different nonnegative constants $\tau$. 
    The authors have studied the problem of inferring the graph structure from a set of observations $\mathbf{y}_k$, each of which corresponds to a different linear combination (i.e., a mixture) of these heat diffusion processes. 
    In contrast, in our problem settings, each observation $\mathbf{y}_k$ corresponds to the output of a single, albeit parametrically uncertain, consensus dynamics.
    For different observations, however, the corresponding consensus dynamics can be identical (Problems~\ref{P:single} and \ref{P:constant_unknown}) or different (Problem~\ref{P:multiple}).

    Comparing the graph filter for consensus dynamics in \eqref{E:consensus_filter} with the graph filter representing heat diffusion, it is easy to show that they have the same asymptotic behavior, i.e., $\lim_{T\to\infty}\prod_{t=1}^T(\mathbf{I}-\alpha_t\mathbf{L})=\lim_{\tau\to\infty}\exp(-\tau\mathbf{L})=\mathbf{11}^{\top}/N$. 
    Indeed, we have $\exp(-\tau\mathbf{L}) = \lim_{j\rightarrow\infty} (\mathbf{I} -\tau\mathbf{L}/j)^j$, and thus when $T$ and $\tau$ take finite values, the graph filters are markedly different.
\end{remark}

In the next two sections we propose different algorithms to solve Problems~\ref{P:single},~\ref{P:constant_unknown} and~\ref{P:multiple}.
Our algorithms are optimization-based and can be considered as two-step procedures.
First, we extract spectral information about the CGL $\mathbf{L}$ from realizations of the output signals, i.e., we collect information about the eigenvectors and eigenvalues of $\mathbf{L}$.
Second, we construct a convex optimization problem to recover the CGL, leveraging the extracted spectral information and incorporating the CGL constraints as well as a sparsity prior.

\section{Topology inference from a single consensus process}\label{S:single}
The focus of this section is on Problem~\ref{P:single}, where we are confronted with a single stationary graph process $h_k(\mathbf{L}) \equiv h(\mathbf{L})$.
Accordingly, it follows from~\eqref{E:general_consensus_filter} and Assumption~\ref{A:input_noise} that
\begin{equation}\label{E:covariance}
	\mathbb{E}[\mathbf{y}_k\mathbf{y}_k^{\top}] = h(\mathbf{L})\mathbb{E}[\bm{\xi}\bm{\xi}^{\top}]h(\mathbf{L})^{\top} = \sigma^2 {h(\mathbf{L})^2},
\end{equation}
where the last equality uses the fact that $\mathbf{L}$ is symmetric.

From~\eqref{E:covariance}, we can see that information about $\mathbf{L}$ is encoded in the covariance of the output $\mathbf{y}$. 
Specifically, as $h(\mathbf{L})=\mathbf{V}h(\mathbf{\Lambda})\mathbf{V}^{\top}$, where $[h(\mathbf{\Lambda})]_{ii}=h(\lambda_i)$ for all $i$, the covariance of the output can be written as
\begin{equation}\label{E:covariance_single_known}
	\mathbf{C}_y=\mathbb{E}[\mathbf{y}_k\mathbf{y}_k^{\top}] = \sigma^2 {h(\mathbf{L})^2} = \sigma^2 \mathbf{V} {h(\mathbf{\Lambda})^2}\mathbf{V}^{\top}.
\end{equation}
The following results are an immediate consequence.

\begin{myproposition}\label{prop:spectrum_cov}
	The {output} covariance matrix $\mathbf{C}_y$ and the graph Laplacian $\mathbf{L}$ share the same set of eigenvectors.
    Moreover, the eigenvalues of $\mathbf{C}_y$ are given by the following transformation of the eigenvalues of $\mathbf{L}$:
\begin{equation}\label{E:eigenvalue_function}
	\lambda_i(\mathbf{C}_y) = \sigma^2 {h(\lambda_i)^2} = \sigma^2 \prod_{t=1}^T (1-\alpha_t\lambda_i)^2.
\end{equation}
\end{myproposition}
\begin{myproof}
    This follows directly from~\eqref{E:covariance_single_known}.
\end{myproof}

\begin{myproposition}
	The input power $\sigma^2$ of the initial condition $\bm{\xi}$ and the eigenvalues of the graph Laplacian $\mathbf{L}$ can be directly recovered from the spectrum of the output covariance $\mathbf{C}_y$.
\end{myproposition}

\begin{myproof}
    Since by assumption $0=\lambda_1<\lambda_2<\cdots<\lambda_N$, it follows from~\cref{prop:spectrum_cov} that $\sigma^2 = \lambda_1(\mathbf{C}_y)>\lambda_2(\mathbf{C}_y)>\cdots>\lambda_N(\mathbf{C}_y)$. 
    Hence, the input power $\sigma^2$ equals the largest eigenvalue of $\mathbf{C}_y$. 

    Furthermore, as $0<\alpha_t< {\lambda_{N}^{-1}}$ for all $t$, we know that $ {\lambda_{N}} <\alpha_{\max}^{-1}$, where $\alpha_{\max}$ is the maximum value of all $\alpha_t$.  
    This implies that $\lambda_i<\alpha_{\max}^{-1}$ for all $i$.
    Defining the scalar function $f(\lambda)=\sigma^2 {h(\lambda)^2}$, it can be seen that $f(\lambda)$ is a monotonically decreasing function of $\lambda$ for $\lambda\leq \alpha_{\max}^{-1}$.
    Hence, the eigenvalues of $\mathbf{L}$ can be uniquely recovered from those of $\mathbf{C}_y$ using~\eqref{E:eigenvalue_function}.
\end{myproof}

Having established that the output covariance $\mathbf{C}_y$ captures all essential information about $\mathbf{L}$, we now discuss how well we can approximate $\mathbf{C}_y$ from a finite number of samples.
Given a set of $M$ independent samples $\{\mathbf{y}_k\}_{k=1}^M$ from an identical distribution with zero-mean and covariance $\mathbf{C}_y$, we can compute the sample covariance matrix 
\begin{equation}\label{E:sample_covariance}
\mathbf{S}_M = \frac{1}{M} \sum_{k=1}^M \mathbf{y}_k\mathbf{y}_k^{\top}
\end{equation}
as an estimate of the covariance $\mathbf{C}_y$. 

The sample covariance $\mathbf{S}_M$ is a real symmetric and positive semi-definite matrix, and we can write its eigendecomposition as $\mathbf{S}_M = \mathbf{U}\mathbf{\Sigma}\mathbf{U}^{\top}$, where the diagonal matrix $\mathbf{\Sigma}$ collects the eigenvalues of $\mathbf{S}_M$ denoted by $\lambda_1(\mathbf{S}_M)\geq \lambda_2(\mathbf{S}_M)\geq \cdots\geq \lambda_N(\mathbf{S}_M)\geq 0$.
The input power $\sigma^2$ can be estimated as 
\begin{equation}\label{E:estimate_sigma}
\hat{\sigma}^2 = \lambda_{\max}(\mathbf{S}_M) = \lambda_1(\mathbf{S}_M).
\end{equation}
We estimate the eigenvectors of $\mathbf{L}$ as $\hat{\mathbf{V}}=\mathbf{U}$ and 
the eigenvalues of $\mathbf{L}$ by replacing $\mathbf{C}_y$ with $\mathbf{S}_M$ in \eqref{E:eigenvalue_function} to obtain
\begin{equation}\label{E:estimate_eigenvalues}
\sqrt{\frac{\lambda_i(\mathbf{S}_M)}{\hat{\sigma}^2}} =h(\hat{\lambda}_i) =\prod_{t=1}^T (1-\alpha_t\hat{\lambda}_i), 
\end{equation}
where $\hat{\lambda}_i$ denotes the estimate of $\lambda_i$.
Notice that in \eqref{E:estimate_eigenvalues} we leverage the fact that $h(\lambda)$ is a non-negative function of $\lambda$ when $\lambda\leq\alpha_{\max}^{-1}$. 
It follows from \eqref{E:estimate_sigma} that $\sqrt{\lambda_i(\mathbf{S}_M)/\hat{\sigma}^2} \in[0,1]$, and $\hat{\lambda}_i$ is the unique real root in the range $[0,\alpha_{\max}^{-1}]$ of the polynomial in \eqref{E:estimate_eigenvalues}. 
After obtaining $\hat{\mathbf{V}}$ and $\hat{\pmb{\lambda}}=[\hat{\lambda}_1,\cdots,\hat{\lambda}_N]^{\top}$, we can estimate $\mathbf{L}$ as $\hat{\mathbf{L}}=\hat{\mathbf{V}}\text{diag}(\hat{\pmb{\lambda}})\hat{\mathbf{V}}^{\top}$.
We refer to this method as \emph{InverseFilter}, and it is summarized in Algorithm~\ref{A:InverseFilter}.
{The same idea has been used in the algorithm proposed in~\cite{Hilmi2018}.}

The theoretical performance guarantees of the InverseFilter approach are stated in Theorems~\ref{T:err_bound_eigenvector} and~\ref{T:err_bound_eigenvalue}.
Both theorems rely on the following instrumental lemma.

\begin{mylemma}\label{L:error_cov}
	Consider $\mathbf{y}_k$ generated as in \eqref{E:general_consensus_filter} for $1\leq k\leq M$ and assume that $M\geq N$. 
	Then for every $\delta>0$, with probability at least $1-\delta$ one has that
	\begin{equation*}
	\|\mathbf{S}_M-\mathbf{C}_y\|_2 \leq C_{\sigma,\delta} \sqrt{N/M},
	\end{equation*}
	where $C_{\sigma,\delta}$ is a constant which depends on $\sigma$ and $\delta$.
\end{mylemma}

\begin{myproof}
	Define a vector $\mathbf{z}\in\mathbb{R}^N$ satisfying $\|\mathbf{z}\|_2=1$. 
	Since $\mathbf{y}_k$ is a zero-mean Gaussian random vector, we have that $\mathbb{E}[\mathbf{z}^{\top}\mathbf{y}_k]=0$ and $\mathbb{E}[e^{\mathbf{z}^{\top}\mathbf{y}_k}]=e^{\frac{1}{2}\mathbf{z}^{\top}\mathbf{C}_y\mathbf{z}}\leq e^{\frac{\sigma^2}{2}}$. 
	Hence, $\mathbf{z}^{\top}\mathbf{y}_k$ is sub-Gaussian satisfying $\mathbb{P}\left[|\mathbf{z}^{\top}\mathbf{y}_k|>t \right]< 2 e^{-\frac{t^2}{2\sigma^2}}$
	\footnote{The concepts of sub-Gaussian random variable and its tail bounds are reviewed in Appendix~\ref{A:random_variable}; see Definition \ref{D:subgaussian} and Lemma \ref{L:subgaussian}.}.  
	The result follows from Proposition 2.1 in \cite{samplecov}. 
\end{myproof}

\begin{mytheorem}[Eigenvectors of InverseFilter]\label{T:err_bound_eigenvector}
	For all $i$, the eigenvectors $\hat{\mathbf{v}}_i$ estimated by InverseFilter satisfy  
	\begin{equation}\label{E:eigenvectors_inversefilter_1}
	\|\hat{\mathbf{v}}_i \!-\! \mathbf{v}_i\|_2 \!\leq\! \frac{2^{3/2}\|\mathbf{S}_M-\mathbf{C}_y\|_2}{\min (\lambda_{i-1}(\mathbf{C}_y)\!-\!\lambda_i(\mathbf{C}_y), \lambda_{i}(\mathbf{C}_y)\!-\!\lambda_{i+1}(\mathbf{C}_y))}, \nonumber
	\end{equation}
	where we set $\lambda_0(\mathbf{C}_y)=\infty$ and $\lambda_{N+1}(\mathbf{C}_y)=-\infty$. Moreover, if $M\geq N$ we have that for every $\delta>0$, with probability at least $1-\delta$,
	\begin{equation}\label{E:eigenvectors_inversefilter_2}
	\|\hat{\mathbf{v}}_i \!-\! \mathbf{v}_i\|_2 \!\leq\! \frac{2^{3/2} C_{\sigma,\delta} \sqrt{N/M}}{\min (\lambda_{i-1}(\mathbf{C}_y)\!-\!\lambda_i(\mathbf{C}_y), \lambda_{i}(\mathbf{C}_y)\!-\!\lambda_{i+1}(\mathbf{C}_y))}.\nonumber
	\end{equation}
\end{mytheorem}

\begin{myproof}
	The first result is a restatement of the Davis-Kahan theorem that follows from Corollary 1 in \cite{dk_variant}. The second result is obtained by combining the first result with the bound in Lemma~\ref{L:error_cov}.
\end{myproof}

Notice that the estimated eigenvectors $\hat{\mathbf{v}}_i$ and the true ones ${\mathbf{v}}_i$ are defined up to a sign inversion. Thus, for the bounds in Theorem~\ref{T:err_bound_eigenvector} to hold both signs have to be picked in a consistent manner. More precisely, we set the sign of our estimated eigenvectors such that $\hat{\mathbf{v}}_i^{\top}\mathbf{v}_i\geq 0$ for all $i$. 

The estimation error on the eigenvalues obtained via InverseFilter can also be bounded, as we show next.

\begin{algorithm}[t]
	\caption{InverseFilter}\label{A:InverseFilter}
	\textbf{Input:} Samples $\{\mathbf{y}_k\}_{k=1}^M$, parameters  $T, \{\alpha_t\}_{t=1}^T$ \\
	\textbf{Output:} Laplacian $\mathbf{L}$
	\begin{algorithmic}[1]
		\State Compute the sample covariance $\mathbf{S}_M$ as in \eqref{E:sample_covariance}
		\State Compute eigendecomposition of  $\mathbf{S}_M=\mathbf{U}\mathbf{\Sigma}\mathbf{U}^{\top}$
		\State Estimate the input power as $\hat{\sigma}^2=\max_{1\leq i\leq N}{\Sigma_{ii}}$
		\State Compute $\{\hat{\lambda}_i\}_{i=1}^N$ by solving \eqref{E:estimate_eigenvalues}
		\State Set $\hat{\mathbf{L}}=\mathbf{U}\mathrm{diag}([\hat{\lambda}_1,\cdots,\hat{\lambda}_N])\mathbf{U}^{\top}$
		\State \textbf{return} $\hat{\mathbf{L}}$
	\end{algorithmic}
\end{algorithm}

\begin{mytheorem}[Eigenvalues of InverseFilter]\label{T:err_bound_eigenvalue}
Assuming that $\sigma$ is known, one has that
\begin{equation}\label{E:eigenvalues_inversefilter_1}
| \hat{\lambda}_i - \lambda_i | < C_{\sigma,\alpha_t,\lambda_i} \|\mathbf{S}_M-\mathbf{C}_y\|_2,
\end{equation}
where $1\leq i\leq N$, $\hat{\lambda}_i$ is obtained using \eqref{E:estimate_eigenvalues} ($\hat{\sigma}$ is set as the ground truth $\sigma$ here), and $C_{\sigma,\alpha_t,\lambda_i}$ is a constant which depends on $\sigma$, $\{\alpha_t\}_{t=1}^T$ and $\lambda_i$. Moreover, if $M\geq N$ we have that for every $\delta>0$, with probability at least $1-\delta$
\begin{equation}\label{E:eigenvalues_inversefilter_2}
| \hat{\lambda}_i - \lambda_i | < C_{\sigma,\alpha_t,\lambda_i, \delta} \sqrt{N/M}.
\end{equation}
\end{mytheorem}

\begin{myproof}
 According to Theorem 6.4.3 in \cite{wielandt_hoffman}, we have that
\begin{equation*}\label{E:bound_eigenvalue_S}
|\lambda_i(\mathbf{S}_M) - \lambda_i(\mathbf{C}_y)| \leq \|\mathbf{S}_M-\mathbf{C}_y\|_2
\end{equation*}
for $1\leq i\leq N$, which implies that
\begin{align}\label{E:bound_square_root}
|\sqrt{\lambda_i(\mathbf{S}_M)} - \sqrt{\lambda_i(\mathbf{C}_y)}| &\leq  \frac{\|\mathbf{S}_M-\mathbf{C}_y\|_2}{\sqrt{\lambda_i(\mathbf{S}_M)} +\sqrt{\lambda_i(\mathbf{C}_y)}} \nonumber \\
 &\leq \frac{\|\mathbf{S}_M -\mathbf{C}_y\|_2}{\sqrt{\lambda_i(\mathbf{C}_y)}}.
\end{align}
It follows from \eqref{E:eigenvalue_function}, \eqref{E:estimate_eigenvalues} and \eqref{E:bound_square_root} that 
\begin{equation}\label{E:bound_h_1}
| h(\hat{\lambda}_i) - h(\lambda_i)| \leq c_{\sigma,\alpha_t,\lambda_i}  \|\mathbf{S}_M-\mathbf{C}_y\|_2,
\end{equation}
where $c_{\sigma,\alpha_t,\lambda_i} = \frac{1}{\sigma\sqrt{\lambda_i(\mathbf{C}_y)}}$ is a constant depending on $\sigma$, $\{\alpha_t\}_{t=1}^T$ and $\lambda_i$.
The derivative of the filter function $h(\lambda)$ is $ h'(\lambda) = -\sum_{k=1}^T \alpha_k \prod_{t\neq k}(1-\alpha_t \lambda)$, and this implies that
\begin{equation*}
\textstyle \min_{\lambda\leq\alpha_{\max}^{-1}} |h'(\lambda)|  \geq \sum_{k=1}^T \alpha_k \prod_{t\neq k}(1-\alpha_t\alpha^{-1}_{\max}) 
 := c_{\alpha}.
\end{equation*}
Since both of $\lambda_i$ and $\hat{\lambda}_i$ are no more than $\alpha_{\max}^{-1}$, we have 
\begin{equation*}
|h(\hat{\lambda}_i)-h(\lambda_i)| >  \min_{\lambda\leq\alpha_{\max}^{-1}} |h'(\lambda)|  \cdot |\hat{\lambda}_i-\lambda_i| \geq c_{\alpha} |\hat{\lambda}_i-\lambda_i|.
\end{equation*}
Combining this with \eqref{E:bound_h_1}, we have that
\begin{equation*}
|\hat{\lambda}_i-\lambda_i| < \frac{c_{\sigma,\alpha_t,\lambda_i}}{ c_{\alpha}} \|\mathbf{S}_M-\mathbf{C}_y\|_2.
\end{equation*}
Set $C_{\sigma,\alpha_t,\lambda_i}={c_{\sigma,\alpha_t,\lambda_i}}/{ c_{\alpha}}$, and the proof of \eqref{E:eigenvalues_inversefilter_1} is completed. Finally, \eqref{E:eigenvalues_inversefilter_2} follows by combining \eqref{E:eigenvalues_inversefilter_1} with the result in Lemma~\ref{L:error_cov}.
\end{myproof}

Theorems~\ref{T:err_bound_eigenvector} and~\ref{T:err_bound_eigenvalue} reveal that, for the InverseFilter method, the estimation error bounds of the eigenvalues and eigenvectors of $\mathbf{L}$ are both proportional to $\|\mathbf{S}_M-\mathbf{C}_y\|_2$, i.e., the estimation error of the covariance $\mathbf{C}_y$.   
Moreover, from Lemma~\ref{L:error_cov} it follows that $\|\mathbf{S}_M-\mathbf{C}_y\|_2$ is bounded by $\sqrt{N/M}$ up to a scalar multiple when $M \!\geq\! N$.
Hence, under the same assumption, the estimation errors of the eigenvalues and eigenvectors of $\mathbf{L}$ obtained by InverseFilter decrease as $1/\sqrt{M}$ with the number of observations.
This implies that InverseFilter is a consistent estimator of the true Laplacian $\bbL$.
In addition, the estimation errors of the eigenvectors depend on the gaps between the eigenvalues of the covariance matrix. 
More precisely, if a given eigenvalue has a large gap with the rest, its corresponding eigenvector can be estimated with high accuracy.

\subsection{Solution to Problem 1: Leverage Laplacian structure}\label{SS:leverage_structure}
One drawback of the InverseFilter method is that the estimate $\hat{\mathbf{L}}$ obtained is generally not a valid CGL.
This is especially a problem when the sample size is small, significantly deteriorating the estimation accuracy.
To alleviate this problem we enforce the constraint that our estimate should (i) be a valid Laplacian belonging to the set $\mathcal{L}_c$ in \eqref{E:CGL}, and (ii) correspond to a sparse graph.
To this end, we solve the following convex optimization problem,
\begin{equation}\label{E:step_2}
	\mathbf{L}^{\ast} = \argmin_{\mathbf{L}} \, d(\mathbf{L},\hat{\mathbf{L}}) + {\beta\|\mathrm{vec}(\mathbf{L})\|_1}, \,\, \text{s.t. } \mathbf{L}\in\mathcal{L}_c,
\end{equation}
where $d(\cdot,\cdot)$ is a convex function which can reflect how close two matrices are, such as {a Bregman divergence}~\cite{bregman_divergences}. 

In this paper, we consider common choices for $d(\mathbf{L},\hat{\mathbf{L}})$ such as $\|\mathbf{L}-\hat{\mathbf{L}}\|_{\mathrm{F}}$ and $\|\mathbf{L}-\hat{\mathbf{L}}\|_{2}$ {as well as their squares}.
In practical problems, the underlying graph is usually sparse, hence we add a regularization term {$\beta\|\mathrm{vec}(\mathbf{L})\|_1$} to promote the graph sparsity, where $\beta$ is a nonnegative regularization parameter.
{The $\ell_1$ norm} is used as a convex surrogate of the non-convex $\ell_0$ pseudo-norm {(i.e., the number of non-zero entries in a vector)}.

The constraint $\mathbf{L}\in\mathcal{L}_c$ in \eqref{E:step_2} leverages the structural information and guarantees that the estimate ${\mathbf{L}}^*$ is a valid CGL.
Note that for $\beta=0$, problem~\eqref{E:step_2} can be interpreted as finding the nearest CGL to $\hat{\mathbf{L}}$.
Hence, we refer to the method recovering $\mathbf{L}^*$ (for general $\beta$) as \emph{NearestCGL}, and summarize it in Algorithm~\ref{A:NearestCGL}.
Finally, notice that although we discuss our problem based on the consensus model in \eqref{E:general_consensus_filter}, the proposed NearestCGL as well as InverseFilter can also be applied to other graph filters whose corresponding functions are nonnegative and one-to-one. 
{
Such functions guarantee that the eigenvalues of $\mathbf{L}$ can be uniquely identified from the eigenvalues of the covariance $\mathbf{C}_y$ [cf. \eqref{E:covariance_single_known}].
}

{
The proposed algorithm can scale to larger networks by translating the optimization problem \eqref{E:step_2} into a well-studied canonical formulation for which efficient solvers exist. 
To this end, let us select $d(\mathbf{L},\hat{\mathbf{L}})=\|\mathbf{L}-\hat{\mathbf{L}}\|_{\mathrm{F}}^2$.
For convenience, we define the vector $\pmb{\ell}:=\mathrm{vec}(\mathbf{L})$ and similarly $\hat{\pmb{\ell}}:=\mathrm{vec}(\hat{\mathbf{L}})$.
We further define two index sets $\mathcal{I}=\{(i-1)N+i \,\,|\,\, 1\leq i\leq N\}$ and $\mathcal{J}=\{(i-1)N+j \,\,|\,\, 1\leq i<j\leq N\}$.
Then $\pmb{\ell}_{\mathcal{I}}$ and $\pmb{\ell}_{\mathcal{J}}$ denote the vectors collecting the diagonal entries and the entries in the lower triangular portion of $\mathbf{L}$, respectively.
We also define a nonnegative vector $\mathbf{a}:=-\pmb{\ell}_{\mathcal{J}}$ which contains the weights of all possible edges as well as $\mathbf{b}:=[\pmb{\ell}_{\mathcal{I}}^{\top}, \sqrt{2}\pmb{\ell}_{\mathcal{J}}^{\top}]^{\top}$ and $\hat{\mathbf{b}}:=[\hat{\pmb{\ell}}_{\mathcal{I}}^{\top}, \sqrt{2}\hat{\pmb{\ell}}_{\mathcal{J}}^{\top}]^{\top}$.
Due to the constraint $\mathbf{L1}=\mathbf{0}$, it is easy to construct a (sparse) matrix $\mathbf{P}$ such that $\mathbf{Pa}=\mathbf{b}$.
It follows that $\|\mathbf{L}-\hat{\mathbf{L}}\|_{\mathrm{F}}^2=\|\mathbf{b}-\hat{\mathbf{b}}\|_2^2=\|\mathbf{Pa}-\hat{\mathbf{b}}\|_2^2$ and $\|\mathrm{vec}(\mathbf{L})\|_1 = 4\|\mathbf{a}\|_1$.
Hence, problem~\eqref{E:step_2} is equivalent to 
\begin{equation}\label{E:fast2}
\mathbf{a}^{\ast} = \argmin_{\mathbf{a}} \|\mathbf{Pa}-\hat{\mathbf{b}}\|_2^2 + 4\beta\|\mathbf{a}\|_1, \text{ s.t. } \mathbf{a}\geq\mathbf{0}.
\end{equation}
Problem \eqref{E:fast2} is in the form of the $\ell_1$-regularized least squares problem with nonnegative constraints.
Efficient methods to solve such problems have been well studied. 
In this paper (Section~\ref{SS:scaling}), we adopt a solver\footnote{The code can be found in~\url{https://web.stanford.edu/~boyd/l1_ls/}.} which uses the truncated Newton interior-point method described in~\cite{fast1}. 
}

\begin{remark}[Solution to Problem~\ref{P:constant_unknown}]\normalfont\label{R:prob_2}
	A potential (naive) solution to Problem~\ref{P:constant_unknown} can be obtained by running multiple instances of the NearestCGL method.
	More precisely, without loss of generality we may absorb the unknown (but constant) $\alpha$ into $\bbL$, thus having only $T$ as an explicit unknown parameter. We may then run NearestCGL for $T \in \{1, 2, \cdots, T_{\max}\}$ for some maximum possible observation time $T_{\max}$, to obtain a series of potential Laplacians $\bbL^*_{(T)}$. We can then choose the best Laplacian among those by selecting the sparsest one or the one that minimizes $d(\bbL^*_{(T)}, {\hat{\mathbf{L}}_{(T)}})$. For the latter case, the intuition is that $\hat{\mathbf{L}}_{(T)}$ is expected to be far from a valid graph Laplacian if a wrong value of $T$ is selected.
\end{remark}
	
	{
\begin{remark}[Related work~\cite{Hilmi2018}] \normalfont
	A related method has been proposed in~\cite{Hilmi2018} that studies a similar problem with graph filters of the form $h_{\tau}(\mathbf{L})=(\mathbf{L}^{\dag})^{\tau}$ where $\tau\in\mathbb{N}$ and $\tau\geq 1$, and the goal is to jointly estimate $\mathbf{L}$ and $\tau$. 
        The authors propose to perform a line search for $\hat{\tau}=1,2,3,\cdots$, and execute their algorithm (as will be explicitly stated in Section~\ref{Ss:num_exp_p1}) for each value of $\hat{\tau}$ to estimate the graph Laplacian. 
        Denote by $\mathbf{L}_{\hat{\tau}}$ the estimate associated with $\hat{\tau}$. 
        For all pairs of $\{\hat{\tau}, \mathbf{L}_{\hat{\tau}}\}$, the minimizer of $\|h_{\hat{\tau}}(\mathbf{L}_{\hat{\tau}})^2-\mathbf{S}_M\|_{\mathrm{F}}$ is then chosen as the solution (assuming unit input power). 

        The minimization of $\|h_{\hat{\tau}}(\mathbf{L}_{\hat{\tau}})^2-\mathbf{S}_M\|_{\mathrm{F}}$ behaves similar to minimizing $\|\mathbf{L}_{\hat{\tau}}-h_{\hat{\tau}}^{-1}(\sqrt{\mathbf{S}_M})\|_{\mathrm{F}}$. 
        As described for the naive solution above, $\mathbf{L}^{\tau/\hat{\tau}}$ is not a valid graph Laplacian, in general, when $\hat{\tau}\neq\tau$.
        Hence, the distance between $\mathbf{L}_{\hat{\tau}}$ (which is a valid graph Laplacian guaranteed by the recovery algorithm) and $h_{\hat{\tau}}^{-1}(\sqrt{\mathbf{S}_M})$ (an estimate of $\mathbf{L}^{\tau/\hat{\tau}}$) is expected to be large for $\hat{\tau}\neq\tau$. However, when the sample size is so small that $h_{\hat{\tau}}^{-1}(\sqrt{\mathbf{S}_M})$ cannot approximate $\mathbf{L}^{\tau/\hat{\tau}}$ well, $h_{\hat{\tau}}^{-1}(\sqrt{\mathbf{S}_M})$ might be close to some graph Laplacian, which is not necessarily the true $\mathbf{L}$. This might lead to a small $\|h_{\hat{\tau}}(\mathbf{L}_{\hat{\tau}})^2-\mathbf{S}_M\|_{\mathrm{F}}$ and, thus, yield a worse solution.

	Due to this issue, we propose a different approach to Problem~\ref{P:constant_unknown}. Instead of estimating the graph Laplacian and the filter parameter at the same time, we adopt a two-step procedure in which we first estimate the unknown parameter and then estimate the graph Laplacian with the parameter estimate. The proposed method is empirically better than the naive one previously described. Since the proposed method relies on our solution to Problem~\ref{P:multiple}, in the next section we tackle Problem~\ref{P:multiple} and defer the solution to Problem~\ref{P:constant_unknown} to Section~\ref{SS:solution_p2}.
\end{remark}}

    \begin{algorithm}[t]
    \caption{NearestCGL (Solution to Problem~\ref{P:single})}\label{A:NearestCGL}
	\textbf{Input:} Samples $\{\mathbf{y}_k\}_{k=1}^M$, parameters $T, \{\alpha_t\}_{t=1}^T, \beta$ \\
	\textbf{Output:} Laplacian $\mathbf{L}$
    \begin{algorithmic}[1]
    \State Run InverseFilter as detailed in Algorithm~\ref{A:InverseFilter}  
    \State Find the nearest sparse CGL ${\mathbf{L}}^*$ to $\hat{\mathbf{L}}$ by solving \eqref{E:step_2}
    \State \textbf{return} ${\mathbf{L}}^*$
    \end{algorithmic}
    \end{algorithm}


\section{Topology inference from multiple consensus processes}\label{S:multiple}

The main focus of this section is on Problem \ref{P:multiple}, where our goal is to infer the network topology $\mathbf{L}\in\mathcal{L}_c$ from a set of observations $\{\mathbf{y}_k\}_{k=1}^M$ each of which corresponds to a different consensus dynamics.
Different $\mathbf{y}_k$ have different covariances $\sigma^2 {h_k(\mathbf{L})^2}$, thus in general the sample covariance $\mathbf{S}_M$ defined in \eqref{E:sample_covariance} will not converge to the covariance of any specific $\mathbf{y}_k$ for increasing sample size $M$.
As a result, the method proposed in Section~\ref{S:single} is not applicable for this setting. 
However, $\mathbf{S}_M$ still contains spectral information about the unknown $\mathbf{L}$ and can be leveraged in the estimation.
We first discuss how information about the eigenvectors of $\mathbf{L}$ can be inferred from $\mathbf{S}_M$.
The result is given in Theorem~\ref{T:eigenvector_converge}, which states that $\mathbf{S}_M$ in \eqref{E:sample_covariance} and $\mathbf{L}$ are simultaneously diagonalizable, provided that the sample size $M$ is sufficiently large. 
For notational purposes, we define the matrix $\mathbf{B}^{(M)}=\mathbf{V}^{\top}\mathbf{S}_M\mathbf{V}$.
The proofs in this section use some properties of random variables which are summarized in Appendix~\ref{A:random_variable}; see Lemmas~\ref{subexp_tail}-\ref{sum_subexp}.

\begin{mytheorem}[Eigenvectors of the sample covariance]\label{T:eigenvector_converge}
For $M\!\to\!\infty$, the eigenbasis $\mathbf{V}$ diagonalizes $\mathbf{S}_M$, i.e., for all $i\neq j$
\begin{equation}\label{E:eigenvector_converge}
    \lim_{M\to\infty} [\mathbf{V}^{\top}\mathbf{S}_M\mathbf{V}]_{ij} = \lim_{M\to\infty} B^{(M)}_{ij} = 0.
\end{equation}
\end{mytheorem}

\begin{myproof}
The matrix $\mathbf{B}^{(M)}$ can be rewritten as
\begin{equation}\label{E:matrix_B}
\mathbf{B}^{(M)} = \frac{1}{M}\sum_{k=1}^M h_k(\mathbf{\Lambda})\mathbf{V}^{\top}\bm{\xi}_k\bm{\xi}_k^{\top}\mathbf{V}h_k(\mathbf{\Lambda})
=\frac{1}{M}\sum_{k=1}^M \mathbf{w}_k\mathbf{w}_k^{\top},
\end{equation}
where we set $\mathbf{w}_k=h_k(\mathbf{\Lambda})\mathbf{V}^{\top}\bm{\xi}_k$.
For each off-diagonal entry in $\mathbf{B}^{(M)}$ we thus have 
\begin{equation*}
B^{(M)}_{ij} = \frac{1}{M}\sum_{k=1}^M [\mathbf{w}_k]_i [\mathbf{w}_k]_j.
\end{equation*}

Since $\bm{\xi}_k\sim\mathrm{N}(\mathbf{0},\sigma^2\mathbf{I})$, it follows from the definition of $\mathbf{w}_k$ that $\mathbf{w}_k\sim\mathrm{N}(\mathbf{0},\sigma^2 {h_k(\mathbf{\Lambda})^2} )$ and thus $[\mathbf{w}_k]_i\sim\mathrm{N}(0,\sigma^2 {h_k(\lambda_i)^2} )$ for all $i$ are independent. 

According to Lemma~\ref{product_gaussian}, $[\mathbf{w}_k]_i [\mathbf{w}_k]_j$ is therefore a sub-exponential random variable with parameters ${(\nu_k,b_k)=(\sqrt{2}\sigma^2h_k(\lambda_i)h_k(\lambda_j),\sqrt{2}\sigma^2h_k(\lambda_i)h_k(\lambda_j))}$.
Hence, $MB^{(M)}_{ij}$ is the sum of $M$ independent zero-mean sub-exponential random variables. 

It follows from Lemma~\ref{sum_subexp} that $MB^{(M)}_{ij}$ is sub-exponential with parameters $(\sqrt{M}\nu_{\ast},b_{\ast})$ where $\nu_{\ast}=\sqrt{\sum_{k=1}^M \nu_k^2/M}$ and $b_{\ast}=\max_{k}b_k$.
Then from Lemma~\ref{subexp_tail}, we have 
\begin{equation*} 
\mathbb{P}\left[ |B_{ij}^{(M)}| \geq l \right] \leq 2\exp\left( -\frac{Ml^2}{2\nu_{\ast}^2} \right)
\end{equation*}
for $0\leq l\leq \nu_{\ast}^2/b_{\ast}$. 
Recall that each $h_{k}(\lambda_i)$ is upper bounded by $1$, hence $\nu_{\ast}^2\leq 2\sigma^4$.
Consequently, for small enough $l>0$:
\begin{equation*}
\lim_{M\to\infty} \mathbb{P}\left[ |B^{(M)}_{ij}| \geq l \right] \leq  \lim_{M\to\infty} 2\exp \left( -\frac{Ml^2}{4\sigma^2} \right) = 0,
\end{equation*}
which completes the proof.
\end{myproof}

Theorem~\ref{T:eigenvector_converge} guarantees that the eigenbasis $\mathbf{V}$ can be recovered by performing the eigendecomposition of $\mathbf{S}_M$ for large enough $M$.
While in most practical instances of network inference $M$ will be bounded, Theorem~\ref{T:eigenvector_converge} can nevertheless be used as a basis for an inference algorithm even if only a finite number of observations are available; see Section~\ref{SS:inference_multiple}.

We remark that the validity of~\eqref{E:eigenvector_converge} does not imply that $\lim_{M\to\infty}\mathbf{S}_M$ exists. 
Indeed, our weak assumptions on the diffusion parameters $T_k$ and $\alpha^{(k)}_{t}$ -- which translate into mild conditions on $h_k(\mathbf{L})$ -- could result in an $\mathbf{S}_M$ that does not converge for increasing $M$.
However, even if $\mathbf{S}_M$ does not converge to a specific matrix, we may interpret \eqref{E:eigenvector_converge} as stating that $\mathbf{S}_M$ converges to the set of matrices diagonalized by $\mathbf{V}$. 
As shown in Theorem~\ref{T:eigenvalue_converage}, $\mathbf{S}_M$ also provides insights about the eigenvalues of the unknown $\mathbf{L}$.


\begin{mytheorem}[Eigenvalues of the sample covariance] \label{T:eigenvalue_converage} Assume that the eigenvalues of $\mathbf{L}=\mathbf{V}\mathbf{\Lambda}\mathbf{V}^{\top}$ are sorted in an increasing order, i.e., $\lambda_1<\lambda_2<\cdots<\lambda_N$. 
For every $\delta>0$ there exists a large enough number of observations $M_{\delta}$ such that the diagonal entries of the matrix $\mathbf{B}^{(M)}=\mathbf{V}^{\top}\mathbf{S}_M\mathbf{V}$ follow an inverse order of the eigenvalues of $\mathbf{L}$, i.e., 
\begin{equation}\label{E:order_Bii}
B^{(M)}_{11} > B^{(M)}_{22} > \cdots > B^{(M)}_{NN},
\end{equation}
with probability at least $1-\delta$ for every $M\geq M_{\delta}$.
\end{mytheorem}

\begin{myproof}
It follows from \eqref{E:matrix_B} that 
\begin{equation*}
B^{(M)}_{ii} = \frac{1}{M} \sum_{k=1}^M( [\mathbf{w}_k]_i)^2,
\end{equation*}
where $[\mathbf{w}_k]_i\sim\mathrm{N}(0,\sigma^2 {h_k(\lambda_i)^2} )$.
According to Lemma~\ref{square_gaussian}, $([\mathbf{w}_k]_i)^2$ is sub-exponential with mean $\sigma^2 {h_k(\lambda_i)^2}$ and parameters $(\nu_k^i,b_k^i)=(2\sigma^2 {h_k(\lambda_i)^2},4\sigma^2 {h_k(\lambda_i)^2} )$.
Thus, we have 
\begin{equation}\label{E:mean_Bii}
	e_i = \mathbb{E}[B^{(M)}_{ii}] = \frac{\sigma^2}{M}\sum_{k=1}^M {h_k(\lambda_i)^2},
\end{equation}
and $MB^{(M)}_{ii}$ is the sum of $M$ independent sub-exponential random variables. 
It follows from Lemma~\ref{sum_subexp} that $MB^{(M)}_{ii}$ is sub-exponential with parameters $(\sqrt{M}\nu_{\ast i },b_{\ast i })$ where $\nu_{\ast i }=\sqrt{\sum_{k=1}^M (\nu_k^{i})^2/M}$ and $b_{\ast i }=\max_k b_k^i$.
Then from Lemma~\ref{subexp_tail}, we have
\begin{equation}\label{E:bound_Bii}
\mathbb{P}\!\left[ |B^{(M)}_{ii}-e_i| \!\geq\! l \right] \!\leq\! 2\exp\!\left(-\frac{Ml^2}{2v_{\ast i}^2}\right)\!\!\leq\! 2\exp\!\left(-\frac{Ml^2}{8\sigma^4}\right)\!,
\end{equation}
for $0\leq l\leq \nu_{\ast i}^2/b_{\ast i}$.
The last inequality follows from the fact that each $h_k(\lambda_i)$ is upper bounded by $1$, which results that $\nu_{\ast i}^2\leq 4\sigma^4$.
A direct application of the union bound on \eqref{E:bound_Bii} yields
\begin{equation*}
\mathbb{P}\left[ \bigcup_{i=1}^N \left|B^{(M)}_{ii}-e_i \right| \geq l\right] \leq 2N\exp\left( -\frac{Ml^2}{8\sigma^4}\right),
\end{equation*}
for $0\leq l\leq \nu_{\ast}^2/b_{\ast}:= \min_{i} (\nu_{\ast i}^2/b_{\ast i})$, from which it immediately follows that
\begin{equation}\label{E:intersection}
\mathbb{P}\left[ \bigcap_{i=1}^N \left|B^{(M)}_{ii}-e_i \right| < l\right] \geq 1- 2N\exp\left( -\frac{Ml^2}{8\sigma^4}\right)\geq 1-\delta,
\end{equation}
where we fixed a desired probability level at $1-\delta$. 
Our goal now is to choose $l$ small enough to ensure that \eqref{E:order_Bii} is satisfied and then solve for the corresponding number of observations $M_{\delta}$ in \eqref{E:intersection} using such an $l$.
To do this, first recall that $1=h_k(\lambda_1)>\cdots>h_k(\lambda_N)$.
We further assume that $h_k(\lambda_i)>h_k(\lambda_j)+\tau$ when $i<j$ for some $\tau>0$, where $\tau$ does not depend on $M$.
It then follows from \eqref{E:mean_Bii} that $e_i>e_j+\sigma^2\tau^2$ for $i<j$.
Consider a deviation from the mean $l^{\ast}:= \sigma^2\tau^2/\gamma$ where $\gamma\geq 2$ is large enough to ensure that $l^{\ast}\leq \nu_{\ast}^2/b_{\ast}$.
By specializing \eqref{E:intersection} to $l=l^{\ast}$ and solving for $M$ as a function of $\delta$, we have that for all $M$ such that
\begin{equation*}
M \geq M_{\delta} := \frac{8 \gamma^2}{\tau^4}\log\left(\frac{2N}{\delta}\right),
\end{equation*}
every random variable $B^{(M)}_{ii}$ is at most a distance $l^{\ast}$ from its mean with probability at least $1-\delta$. 
Since by definition $l^{\ast}<(e_i-e_j)/2$ for $i<j$, the variables $B^{(M)}_{ii}$ are sorted in the same order as their means with high probability, and the proof is completed.
\end{myproof}

Theorem~\ref{T:eigenvalue_converage} reveals that, no matter whether $\mathbf{S}_M$ converges to a specific matrix or not, the diagonal entries of $\mathbf{V}^{\top}\mathbf{S}_M\mathbf{V}$ follow a specific order, i.e., the inverse order of the eigenvalues of the true CGL $\mathbf{L}$ with high probability. 
This observation, in combination with Theorem~\ref{T:eigenvector_converge}, is leveraged in Section \ref{SS:inference_multiple} to develop a network topology inference algorithm for finite $M$.


\subsection{Solution to Problem 3: Leverage spectral ordering}\label{SS:inference_multiple}

The recovery of the CGL $\mathbf{L}$ under the setting of Problem \ref{P:multiple} is generally an underdetermined problem.
Even when we fix the true eigenbasis $\mathbf{V}$ and the ordering of the eigenvalues, there still exists freedom in choosing the exact eigenvalues as long as the order is preserved.
To sort out this ambiguity, which amounts to selecting the eigenvalues following a specific order, we assume that the network topology of interest is optimal in some sense.
Following the criterion considered in NearestCGL to solve Problem~\ref{P:single} [cf.~\eqref{E:step_2}], {in this paper we focus on the optimality based on sparsity. Notice that alternative criteria can be introduced in the form of a generic convex function by replacing the objective in \eqref{objective}.} 

Recall that we denote by $\mathbf{S}_M=\mathbf{U}\mathbf{\Sigma}\mathbf{U}^{\top}$ the eigendecomposition of the sample covariance defined in \eqref{E:sample_covariance} where the eigenvalues in $\mathbf{\Sigma}$ are sorted in decreasing order.
Our inferred CGL ${\mathbf{L}}^*$ can be found by solving the following convex optimization problem.
\begin{subequations}\label{E:solution_p3}
	\begin{align}
			\{ \mathbf{L}^{\ast}, \,& \pmb{\gamma}^{\ast} \} = \argmin_{\{\mathbf{L},\,\pmb{\gamma}\}} \|\mathrm{vec}(\mathbf{L})\|_1  \label{objective}\\
 \text{s.t. }\,\, & \mathbf{L}\in\mathcal{L}_c, \label{constraint1}\\
 & d(\mathbf{L}, \mathbf{U}\diag(\pmb{\gamma})\mathbf{U}^{\top}) \leq \epsilon, \label{constraint2}\\
 & \gamma_N = 1, \,\, \gamma_i\leq \gamma_{i+\eta} \text{ for } i=1,\cdots,N-\eta. \label{constraint3} 
\end{align}
\end{subequations}
Analogously, to the discussion after \eqref{E:step_2}, {the $\ell_1$ norm} is a convex relaxation of the $\ell_0$ pseudo-norm, thus the objective \eqref{objective} promotes sparsity in the estimate ${\mathbf{L}}^{\ast}$. Alternatively, {$\|\mathrm{vec}(\mathbf{L})\|_1$} can be replaced by its iterative reweighted counterpart~\cite{reweighted}, which has shown to perform better in practice.

 \begin{algorithm}[t]
	\caption{OrderedSpecTemp (Solution to Problem~\ref{P:multiple})}\label{A:orderedspectemp}
	\textbf{Input:} Samples $\{\mathbf{y}_k\}_{k=1}^M$, $\epsilon$, $\eta$ \\
	\textbf{Output:} Laplacian $\mathbf{L}$
	\begin{algorithmic}[1]
		\State Compute the sample covariance {as in \eqref{E:sample_covariance}}
		\State Obtain $\mathbf{U}$ by eigendecomposition $\mathbf{S}_M=\mathbf{U}\mathbf{\Sigma}\mathbf{U}^{\top}$ where the eigenvalues {in} $\mathbf{\Sigma}$ are sorted in decreasing order
		\State Solve problem~\eqref{E:solution_p3}
		\State \textbf{return} ${\mathbf{L}}^*$
	\end{algorithmic}
\end{algorithm}

The constraint in \eqref{constraint1} forces ${\mathbf{L}}^*$ to be a valid CGL. 
The constraints in \eqref{constraint2} impose that ${\mathbf{L}}^*$ must be close to being diagonalized by $\mathbf{U}$, i.e. the eigenbasis of the sample covariance $\mathbf{S}_M$.
It has been shown in Theorem~\ref{T:eigenvector_converge} that $\mathbf{U}$ coincides with $\mathbf{V}$, i.e. the eigenbasis of $\mathbf{L}$, for arbitrarily large sample size $M$.
However, for all practical implementations, $M$ is finite and thus, we do not require ${\mathbf{L}}^*$ to be diagonalized by $\mathbf{U}$ directly. 
Rather, we require ${\mathbf{L}}^*$ to be close to another matrix which is diagonalized by $\mathbf{U}$.
The matrix distance is measured by the convex function $d(\cdot,\cdot)$.
The problem can be reduced to a linear program if we choose {$d(\cdot,\cdot)$ as the maximum norm.}
Lastly, the constraints in \eqref{constraint3} leverage Theorem~\ref{T:eigenvalue_converage} and incorporate the fact that the eigenvalues of $\mathbf{S}_M$ and the true CGL $\mathbf{L}$ are inversely ordered by forcing {the entries in $\pmb{\gamma}$} to follow an approximately increasing order.
The parameter $\eta$ is a positive integer which can be adjusted according to the sample size $M$.
When $\eta=1$ we impose a strict order on the eigenvalues for the cases in which $M$ is sufficiently large whereas for $\eta>1$ we impose a laxer order for the cases where $M$ is not large enough.
We set $\gamma_N=1$ to avoid the trivial solution ${\mathbf{L}}^*=\mathbf{0}$.
We can also set $\gamma_N$ to other positive values since it will only vary the scale of the estimate ${\mathbf{L}}^*$.
Notice that this scale ambiguity is insurmountable given that a common factor across all unknown diffusion rates $\alpha^{(k)}_t$ can be absorbed into the unknown $\mathbf{L}$.
We denote the proposed method as \emph{OrderedSpecTemp} given that it uses sorted eigenvectors (spectral templates) in the recovery of $\bbL^*$, and we summarize it in Algorithm~\ref{A:orderedspectemp}.



Notice that in~\eqref{E:solution_p3} we may move the constraint~\eqref{constraint2} to the objective function and utilize the $\ell_1$ norm as the regularization term. 
The objective function then becomes $d(\mathbf{L}, \mathbf{U}\mathrm{diag}(\pmb{\gamma})\mathbf{U}^{\top}) + \beta\|\mathrm{vec}(\mathbf{L})\|_1$, where the graph sparsity can be tuned with the regularization parameter $\beta$ in addition to the spectral fitting term. 
As is the case for existing work~\cite{smooth0,smooth1,Hilmi2018,heat_diffusion}, the accuracy of prior knowledge, i.e., the graph sparsity level, will influence the quality of the selected regularization parameter and thus affect the algorithm performance.

	It should also be noticed that in Algorithm~\ref{A:orderedspectemp} we are able to recover $\mathbf{L}$ without the need of recovering the set of parameters $\alpha_{t}^{(k)}$.
	Indeed, under the setting of Problem~\ref{P:multiple}, $\alpha_{t}^{(k)}$ can be different over $k$ and different over $t$.
This weak assumption makes it challenging (if not impossible) to identify $\{\alpha_{t}^{(k)}\}$. Even for the simplified case in which $\alpha_t^{(k)}$ is identical for every $k$ but potentially varying with $t$ (i.e., we consider multiple snapshots of an identical linear time-varying dynamics), jointly estimating $\mathbf{L}$ and $\{\alpha_t\}$ is challenging. We leave this research direction for future work.

	We want to highlight that, while Algorithm \ref{A:orderedspectemp} is proposed under the setting of Problem \ref{P:multiple}, it can also be applied to unknown filters of the form $h_k(\mathbf{L})\equiv h(\mathbf{L})$. The only requirement is that the filter $h(\lambda)$ is a nonnegative and decreasing function of $\lambda$, which corresponds to a certain type of low-pass graph filter.
Under this condition, the eigenvalues of the covariance $\mathbf{C}_y$ follow the inverse order of the eigenvalues of $\mathbf{L}$ (if we fix the order of the eigenvectors). In this case, the constraint incorporating the eigenvalue order information is still valid. This constraint is not as restrictive as it might look at first sight. 
For example, heat diffusion processes~\cite{heat,heat_diffusion} satisfy this condition.

Finally, let us outline a way of solving problem~\eqref{E:solution_p3} efficiently.
We select $d(\cdot,\cdot)$ as the squared Frobenius norm.
Note that we have $\|\mathbf{L}-\mathbf{U}\diag(\pmb{\gamma})\mathbf{U}^{\top}\|_{\mathrm{F}}^2=\|\mathrm{vec}(\mathbf{L})-(\mathbf{U}\odot\mathbf{U})\pmb{\gamma}\|_2^2$ where $\odot$ denotes the Khatri-Rao product.
Hence problem~\eqref{E:solution_p3} can also be formulated as
\begin{subequations}\label{E:fast_p3}
\begin{align}
\{ \mathbf{L}^{\ast}, & \pmb{\gamma}^{\ast} \}  = \argmin_{\{\mathbf{L},\pmb{\gamma}\}} \|\mathrm{vec}(\mathbf{L}) - (\mathbf{U}\odot\mathbf{U})\pmb{\gamma}\|_2^2 + \beta\|\mathrm{vec}(\mathbf{L})\|_1  \\
 \text{s.t. }\,\, & \mathbf{L}\in\mathcal{L}_c, \\
 & \gamma_N = 1, \,\, \gamma_i\leq \gamma_{i+\eta} \text{ for } i=1,\cdots,N-\eta,
\end{align}
\end{subequations}
where we have moved the constraint \eqref{constraint2} to the objective function and used the $\ell_1$ norm as the regularizer.
We can solve the convex problem~\eqref{E:fast_p3} via block coordinate descent by alternately optimizing $\mathbf{L}$ and $\pmb{\gamma}$.
If we fix $\pmb{\gamma}$ and optimize $\mathbf{L}$, the problem will be the same as \eqref{E:fast2}.
If we fix $\mathbf{L}$ and optimize $\pmb{\gamma}$, the problem is in fact in the form of a least-squares minimization with linear inequality constraints and, again, standard solver packages can be used. 

\begin{remark}[{Related work \cite{Segarra2016}}] \normalfont
Leveraging the ordering information of the eigenvectors -- in terms of the associated eigenvalues -- is essential to the performance of OrderedSpecTemp. Indeed, in \cite{Segarra2016}, a method is proposed based on the spectral templates without the order information to recover adjacency and normalized Laplacian matrices from diffused signals.
Such a method cannot be directly extended to our case in order to recover CGL matrices.
Notice that we consider a different graph shift operator compared to~\cite{Segarra2016}.
If we do not leverage the order information, i.e. ignoring \eqref{constraint3}, and add one more constraint such as $\tr(\mathbf{L})\geq 1$ to problem \eqref{E:solution_p3} to avoid the solution ${\mathbf{L}}^*=\mathbf{0}$, and consider the infinite sample size case in which we have exact spectral templates, then the solution to \eqref{E:solution_p3} will be uninformative. More precisely, we would recover ${L}^*_{ii}=\frac{1}{N}$ and ${L}^*_{ij}=-\frac{1}{N(N-1)}$ for $i\neq j$.
In addition, \cite{Segarra2016} considers a stationary graph process in which multiple observations are obtained from an identical dynamics, but in Problem~\ref{P:multiple} we consider a more practical and challenging case where observations are sampled from different dynamics.
\end{remark}

    \subsection{Solution to Problem 2: Combine NearestCGL and OrderedSpecTemp}\label{SS:solution_p2}

Thus far we have proposed solutions to Problem~\ref{P:single} (namely, NearestCGL) and Problem~\ref{P:multiple} (namely, OrderedSpecTemp).
Based on these, we now propose a solution to Problem~\ref{P:constant_unknown} that empirically outperforms the one delineated in Remark~\ref{R:prob_2}.

First notice that for Problem~\ref{P:constant_unknown}, $\mathbf{L}$ can only be recovered up to a scalar multiple since $\alpha$ is unknown and $\alpha\mathbf{L} = \frac{\alpha}{\alpha_0}(\alpha_0\mathbf{L})$ for any non-zero scalar $\alpha_0$.
Since $T$ is unknown, NearestCGL as stated in Algorithm~\ref{A:NearestCGL} cannot be applied. More precisely, \eqref{E:estimate_eigenvalues} cannot be solved to find the estimated eigenvalues in InverseFilter.
However, we may still apply OrderedSpecTemp as explained in Algorithm~\ref{A:orderedspectemp}, and we denote its solution as ${\mathbf{L}}^*_{\mathrm{ord}}$.
To estimate the unknown observation time $T$, we leverage the fact that $T$ is an integer and perform a line search to optimize
\begin{equation}\label{E:T_est}
\hat{T} = \argmin_{t \in \{1, \cdots, T_{\max}\}}\,\, \|{\mathbf{L}}^*_{\mathrm{ord}} - \hat{\mathbf{L}}_t  \, \tr({\mathbf{L}}^*_{\mathrm{ord}})/\tr(\hat{\mathbf{L}}_t)\|_{\mathrm{F}},
\end{equation}
where $\hat{\mathbf{L}}_t$ is obtained via InverseFilter (cf. Algorithm~\ref{A:InverseFilter}) by assuming that $T=t$. 
{The ratio of traces in \eqref{E:T_est} ensures that the scales of ${\mathbf{L}}^*_{\mathrm{ord}}$ and $\hat{\mathbf{L}}_t$ are comparable.}
Notice that for this problem setting where the diffusion rates $\alpha_t \equiv \alpha$ are constant, InverseFilter can be simplified as
\begin{equation*}
\hat{\mathbf{L}}_t = \mathbf{I} - (\mathbf{S}_M/\hat{\sigma}^2)^{\frac{1}{2t}},
\end{equation*}
where $\hat{\sigma}^2$ is obtained via \eqref{E:estimate_sigma}. 
Intuitively, if ${\mathbf{L}}^*_{\mathrm{ord}}$ were to coincide with the true Laplacian $\bbL$, then we would expect $\hat{T}$ in \eqref{E:T_est} to return the true observation time $T$. 
In the absence of such ground truth, OrderedSpecTemp provides an estimate of the Laplacian that relies on the available observation. 
However, Algorithm~\ref{A:orderedspectemp} does not leverage the fact that all the observations come from a single consensus dynamics with a constant diffusion rate (cf. Problem~\ref{P:constant_unknown}). 
Hence, once we have estimated $T$, we take advantage of this fact by implementing NearestCGL (cf. Algorithm~\ref{A:NearestCGL}) for $T = \hat{T}$, to obtain our estimated Laplacian $\bbL^*$.
This hybrid algorithm is summarized in Algorithm~\ref{A:hybrid}.

 \begin{algorithm}[t]
	\caption{Hybrid (Solution to Problem~\ref{P:constant_unknown})}\label{A:hybrid}
	\textbf{Input:} Samples $\{\mathbf{y}_k\}_{k=1}^M$, $\epsilon$, $\eta$, $\beta$ \\
	\textbf{Output:} Laplacian $\mathbf{L}$
	\begin{algorithmic}[1]
		\State Run OrderedSpecTemp (Algorithm~\ref{A:orderedspectemp}) to obtain ${\mathbf{L}}^*_{\mathrm{ord}}$ 
		\State Solve \eqref{E:T_est} to obtain $\hat{T}$
		\State Run NearestCGL (Algorithm~\ref{A:NearestCGL}) {with $\hat{T}$} to obtain ${\mathbf{L}}^*$
		\State \textbf{return} ${\mathbf{L}}^*$
	\end{algorithmic}
\end{algorithm}

\begin{remark}\normalfont
It should be noted that the assumption $\alpha_t \equiv \alpha$ is made only in the context of Problem~\ref{P:constant_unknown}.
For such dynamics whose diffusion rates do not change over time, we can assume that $\alpha_t$ is constant and adopt Algorithm~\ref{A:hybrid}.
When we do not have such prior information, we can leverage Algorithm~\ref{A:orderedspectemp} which imposes less assumptions on the diffusion parameters. 
Intuitively, in Problems~\ref{P:single}-\ref{P:multiple}, we assume different levels of knowledge about $\{\alpha^{(k)}_t\}$ to find a trade-off between model flexibility and algorithm performance.
In fact, we can use Algorithm~\ref{A:orderedspectemp} to solve all of the three problems.
However, if we \emph{do have} additional prior information about the model (like in Problems~\ref{P:single} and \ref{P:constant_unknown}), we can incorporate it into our algorithm to further improve the estimation accuracy.
This point is validated in our experiments; see Fig.~\ref{Fig:T_est_err}.
\end{remark}

\section{Numerical experiments}\label{S:num_exp}

We illustrate the performance of the methods described and compare them with state-of-the-art solutions through a series of numerical experiments. 
The experiments based on synthetic data are divided by the problem type that is being studied (Sections~\ref{Ss:num_exp_p1}-\ref{Ss:num_exp_p3}) whereas in Section~\ref{Ss:num_exp_real} we implement our methods to real-world data.
{In addition, we discuss the scalability of the proposed algorithms and infer a larger real-world network in Section~\ref{SS:scaling}.}

\subsection{Experiments on Problem~\ref{P:single}}\label{Ss:num_exp_p1}

\begin{figure*}
	\centering
	\subfigure[]{
			\centering
			\includegraphics[width=0.3\textwidth]{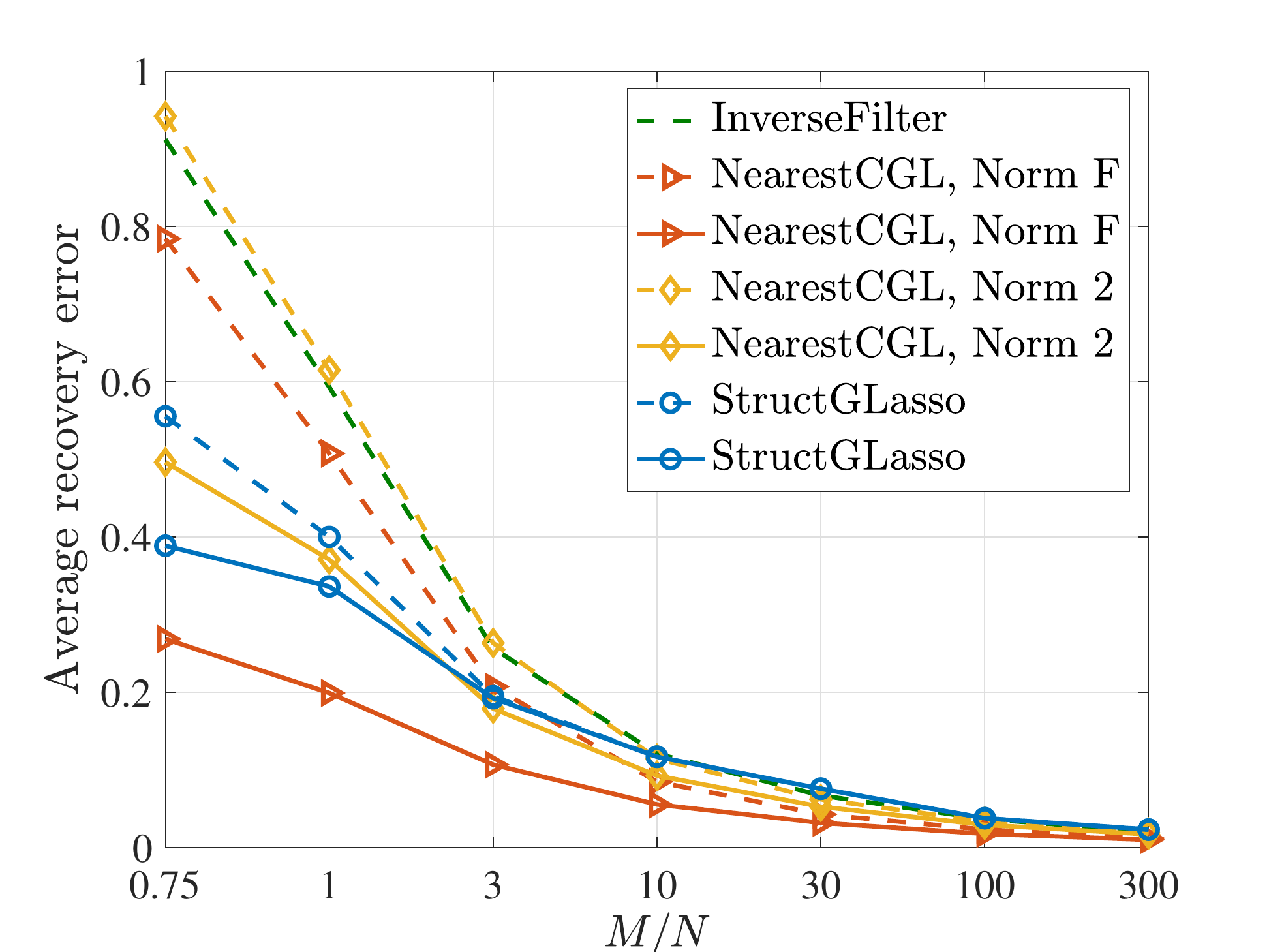}
			\label{Fig:ER_Gaussian}
		}	
	\subfigure[]{
			\centering
			\includegraphics[width=0.3\textwidth]{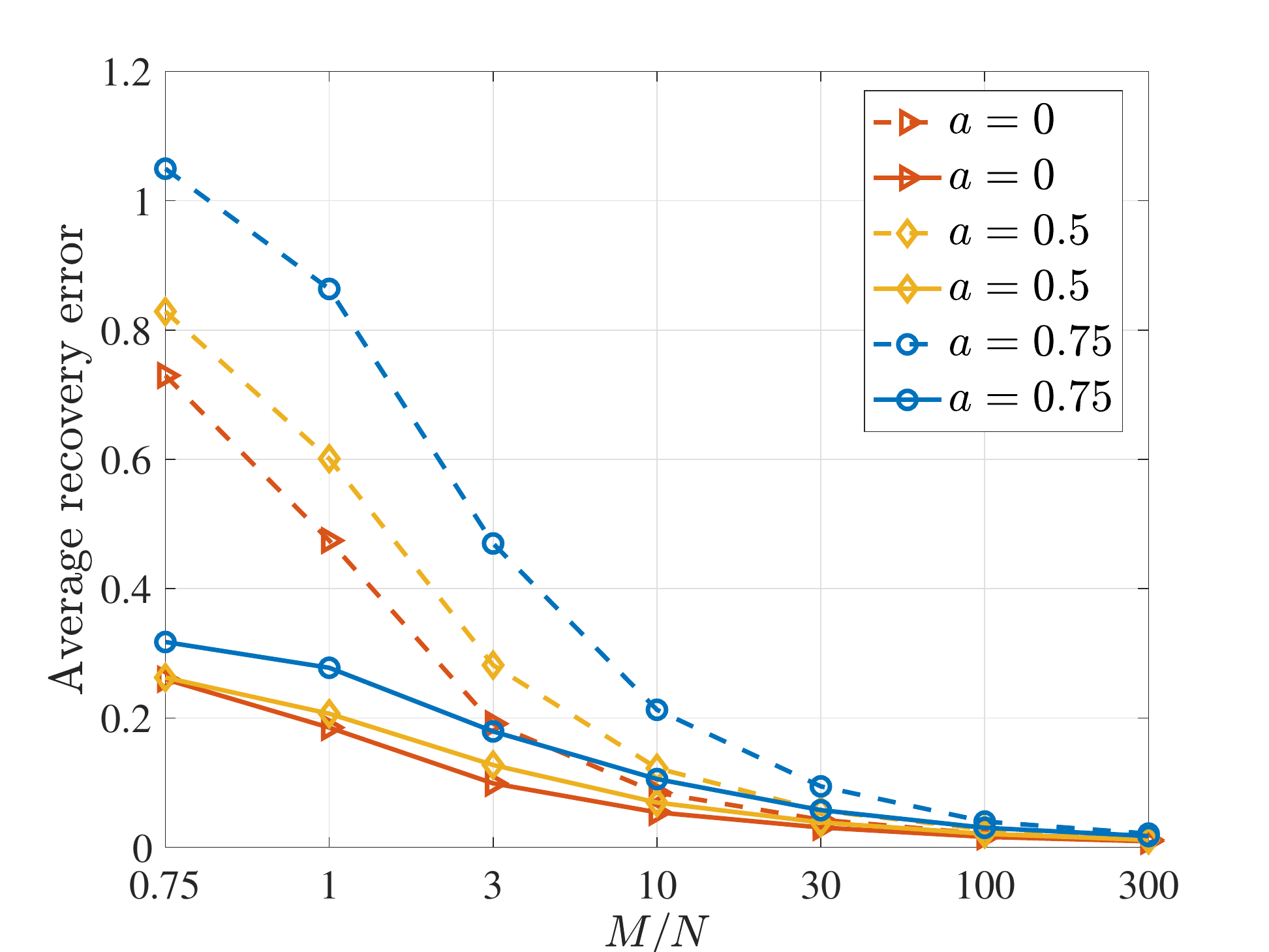}
			\label{Fig:time}
		}	
	\subfigure[]{
			\centering
			\includegraphics[width=0.3\textwidth]{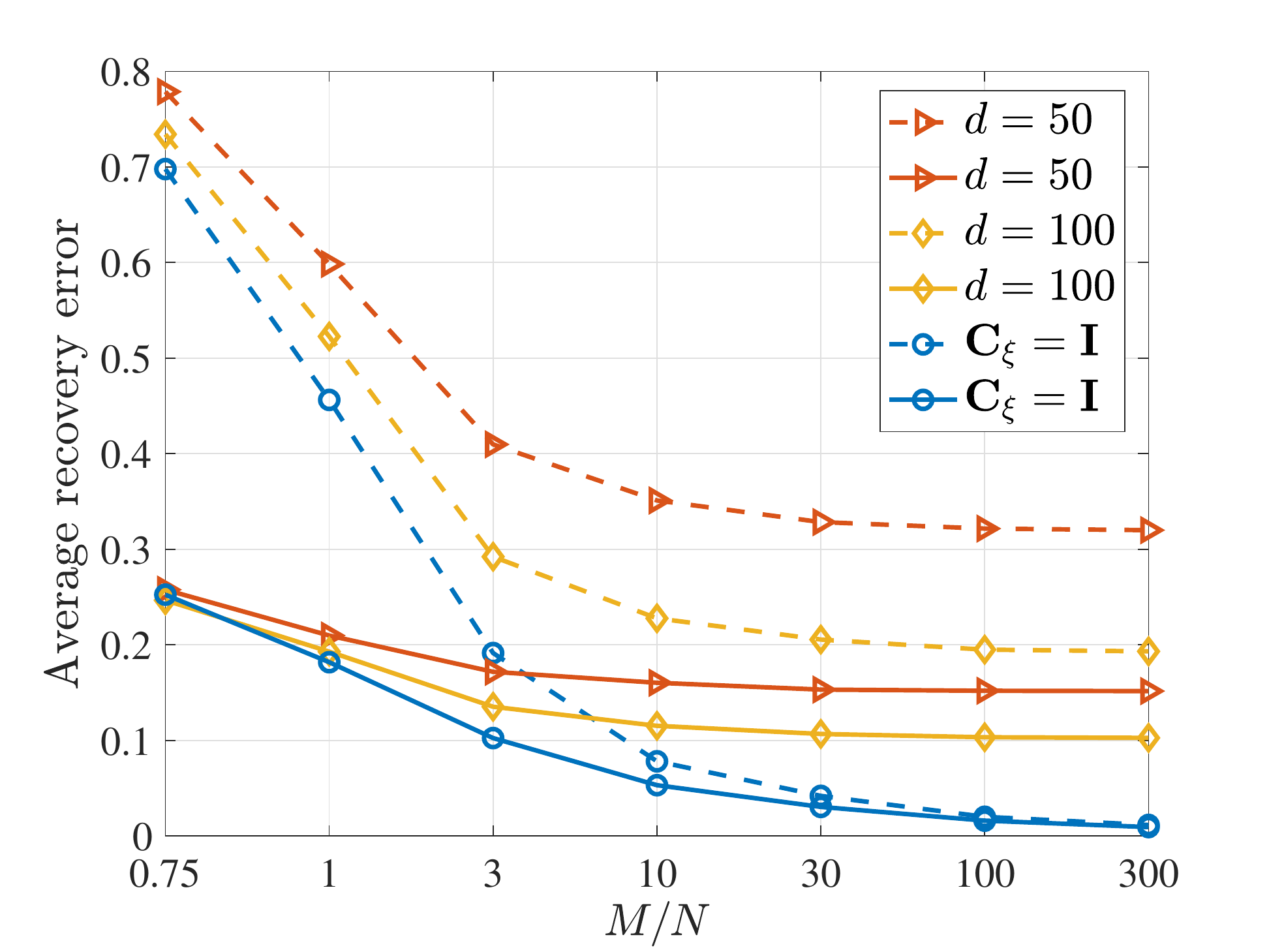}
			\label{Fig:wishart}
		}
		\caption{\small {Average recovery error as a function of the number of observations under the setting of Problem~\ref{P:single} for {ER graphs}. Solid (dashed) lines correspond to methods with (without) regularization. (a) Performance comparison between InverseFilter (Algorithm~\ref{A:InverseFilter}), NearestCGL (Algorithm~\ref{A:NearestCGL}) for two types of distances, and StructGLasso~\eqref{E:graphical_lasso} when white Gaussian inputs are considered. (b) Performance of NearestCGL with Frobenius {norm} when the input is generated as time series with different correlation parameters $a$. 
		(c) Performance of NearestCGL with Frobenius {norm} when the input covariance follows the normalized Wishart distribution with different degrees of freedom~$d$.} }
\end{figure*}

We compare the performance of InverseFilter (Algorithm~\ref{A:InverseFilter}), NearestCGL (Algorithm~\ref{A:NearestCGL}), and the approach proposed in \cite{Hilmi2018} for estimating the network topology from a single consensus dynamics.
The method in \cite{Hilmi2018} assumes that the graph filter is a one-to-one and non-negative function of the CGL $\mathbf{L}$ with one unknown parameter, and proposes an approach to jointly estimate $\mathbf{L}$ and the unknown parameter from filtered signals.
We refer to this approach as StructGLasso in this paper\footnote{We thank the authors of~\cite{Hilmi2018} for kindly sharing their code with us.}, and it can be directly adopted for our problem setting where the parameters in the graph filter are given.  
The StructGLasso method solves the following problem to estimate $\mathbf{L}$,
\begin{equation}\label{E:graphical_lasso}
	{\mathbf{L}}^*_{\mathrm{sgl}} = \argmin_{\mathbf{L}} \, \tr(\mathbf{L}\hat{\mathbf{L}}^{\dag}) - \log |\mathbf{L}| + \beta{\|\mathrm{vec}(\mathbf{L})\|_1}, \text{ s.t. } \mathbf{L}\in\mathcal{L}_c,
\end{equation}
where $\hat{\mathbf{L}}$ is the estimate obtained via InverseFilter, and $|\mathbf{L}|$ denotes the pseudo-determinant of $\mathbf{L}$.
Notice that the idea behind \eqref{E:graphical_lasso} is to incorporate structural information into graphical LASSO \cite{GLasso0, Hilmi2017}.

We generate synthetic datasets based on the signal model in Problem~\ref{P:single}.
We consider random Erd\H{o}s-R\'{e}nyi (ER) graphs \cite{random_graphs} of size $N=36$ and edge-formation probability $p=0.1$.
The edge weights are randomly selected from a uniform distribution in the interval $(0.1,3)$. 
We consider white Gaussian input, i.e. $\bm{\xi}\sim\mathrm{N}(\mathbf{0},\sigma^2\mathbf{I})$, and we set $\sigma=1$.
Therefore, the dataset entries $\{\mathbf{y}_k\}_{k=1}^M$ are randomly drawn from the distribution $\mathrm{N}(\mathbf{0}, {h(\mathbf{L})^2} )$ where $h(\mathbf{L})=\prod_{t=1}^T(\mathbf{I}-\alpha_t\mathbf{L})$. 
We set $T=3$ and $\{\alpha_1,\alpha_2,\alpha_3\}=\{0.7,0.8,0.9\}/\lambda_{\max}(\mathbf{L})$. 
To measure the performance, we compute the recovery error as $\|{\mathbf{L}}^* - \mathbf{L}\|_{\mathrm{F}}/\|\mathbf{L}\|_{\mathrm{F}}$ where $\mathbf{L}$ is the ground truth and ${\mathbf{L}}^*$ is the estimate. 
For NearestCGL and StructGLasso, we consider two cases where (i) there is no regularization ($\beta = 0$), and (ii) the regularization parameter $\beta$ is carefully selected.  
For StructGLasso, $\beta$ is chosen from the following set 
$$\{0\}\cup\{0.75^r s_{\max}\sqrt{\log(N)/M} \,|\, r=1,2,\cdots,14\}$$ 
as suggested in \cite{Hilmi2017,Hilmi2018} where $s_{\max}=\max_{i\neq j}|[\hat{\mathbf{L}}^{\dag}]_{ij}|$ is the maximum off-diagonal entry of $\hat{\mathbf{L}}^{\dag}$ in absolute value.  
For NearestCGL, we consider two choices for $d(\mathbf{L},\hat{\mathbf{L}})$ in \eqref{E:step_2}, namely $\|\mathbf{L}-\hat{\mathbf{L}}\|_{\mathrm{F}}$ and $\|\mathbf{L}-\hat{\mathbf{L}}\|_{2}$.
For NearestCGL with $\|\cdot\|_{\mathrm{F}}$, $\beta$ is selected from $\{0,0.055\!:\!0.0025\!:\!0.085\}$ for $M/N\leq3$ and from $\{0\!:\!0.01\!:\!0.08\}$ otherwise.
For NearestCGL with $\|\cdot\|_{2}$, $\beta$ is selected from $\{0,0.008\!:\!0.001\!:\!0.02\}$ for $M/N\leq3$ and from $\{0\!:\!0.002\!:\!0.016\}$ otherwise.
We perform Monte-Carlo simulations and compute the average recovery error over $20$ realizations. 
The methods are all implemented using CVX~\cite{cvx} and the results are shown in Fig.~\ref{Fig:ER_Gaussian}.

It can be observed that, for small sample size, the regularization improves the estimation performance significantly. 
For NearestCGL, the matrix distance based on the Frobenius norm outperforms {the one} based on the {spectral norm}.
Moreover, the proposed NearestCGL with Frobenius norm and a fine tuned regularization parameter outperforms all of the other methods for every sample size considered.

\begin{remark}[Related work~\cite{Hilmi2018}] \normalfont \label{R:comp_hilmi}
	Both our proposed NearestCGL and the algorithm in \cite{Hilmi2018} can be applied to the class of graph filters whose corresponding functions are nonnegative and one-to-one if the type and parameters of the graph filter are known. 
	The first step of these two algorithms are similar (i.e., InverseFilter) while their second steps differ significantly.
	It can be observed from \eqref{E:step_2} and \eqref{E:graphical_lasso} that our second step directly relies on $\hat{\mathbf{L}}$ while \eqref{E:graphical_lasso} is based on the pseudo-inverse $\hat{\mathbf{L}}^{\dag}$.
	Compared to \cite{Hilmi2018}, we use different functions $d(\cdot,\cdot)$ to measure the difference between two matrices.
    We adopt matrix norms, whereas~\cite{Hilmi2018} uses a log-determinant Bregman divergence~\cite{bregman_divergences}.
	The specific graph filter type will influence the estimation accuracies of $\mathbf{L}$ and $\mathbf{L}^{\dag}$ obtained in the first step, which causes the performance differences between our algorithm and the one in \cite{Hilmi2018}.
	Roughly speaking, \cite{Hilmi2018} shows the advantage of its proposed algorithm for a few types of graph filters which explicitly involve the pseudo-inverse of $\mathbf{L}$.
	This is because, for such graph filters, the estimation accuracy of $\mathbf{L}^{\dag}$ obtained by InverseFilter is generally higher than that of $\mathbf{L}$.
	One the other hand, our model can be extended beyond consensus to a wider range of graph filters that are explicit functions of $\mathbf{L}$ including heat diffusion $h(\mathbf{L})=\exp(-\tau\mathbf{L})$. In this respect, both approaches seem to complement each other since they work better for different classes of filters.
\end{remark}

For conciseness we only present the results for white Gaussian input and ER graphs, while the observed results are preserved when considering white uniform input -- i.e., each entry in the input $\bm{\xi}$ is randomly generated from a uniform distribution in the interval $[-\sqrt{3}\sigma,\sqrt{3}\sigma]$ -- and other graph models including grid graphs, stochastic block model (SBM), small-world, and Barab\'{a}si-Albert model \cite{random_graphs}.

To study the influence in the performance of a deviation from the independent, white Gaussian input assumption (Assumption~\ref{A:input_noise}), we consider two types of inputs: 
(i) we generate the input as a time series of the form $\bm{\xi}_t =a \bm{\xi}_{t-1} +(1-a)\mathbf{w}_t$ where $\bm{\xi}_1$ and $\{\mathbf{w}_t\}$ are randomly drawn from $\mathrm{N}(\mathbf{0},\mathbf{I})$; and 
(ii) we generate the input following the distribution $\mathrm{N}(\mathbf{0},\mathbf{C}_{\pmb{\xi}})$ where $\mathbf{C}_{\pmb{\xi}}$ is drawn from the Wishart distribution $\mathrm{W}_N(\mathbf{I},d)$ and normalized by $d$.
The other simulation parameters are the same as those used in generating Fig.~\ref{Fig:ER_Gaussian}.
The results are shown in Figs.~\ref{Fig:time} and~\ref{Fig:wishart}, respectively.

It can be observed that, for case (i), the performance of NearestCGL decreases as the parameter $a$ increases.
This is as expected since larger values of $a$ introduce more correlation between successive inputs, thus deviating from the independent input assumption ($a=0$).
However, regardless of the value of $a$, the recovery error decreases as the sample size $M$ increases.
Intuitively, this can be attributed to the fact that $\bm{\xi}_{t_1}$ and $\bm{\xi}_{t_2}$ are close to independent if they are separated by a long time period.
Hence, NearestCGL can be seen to output a consistent estimator under this setting.

For case (ii), the recovery error of NearestCGL decreases as the parameter $d$ increases. 
This is also natural since the resulting $\mathbf{C}_{\pmb{\xi}}$ is closer to the identity matrix for larger $d$, and the white case $\mathbf{C}_{\pmb{\xi}}=\mathbf{I}$ can be considered as the extreme setting when $d=\infty$.
For finite $d$, the recovery error does not converge to zero for large $M$.
This follows from \eqref{E:covariance}, where a colored input implies that $\bbC_y$ does not share the same set of eigenvectors as $\bbL$. 
Hence, NearestCGL does not output a consistent estimator in this setting.

\subsection{Experiments on Problem~\ref{P:constant_unknown}}\label{Ss:num_exp_p2}


\begin{figure}
	\centering
	\subfigure[]{
		\centering
		\includegraphics[width=0.4\textwidth]{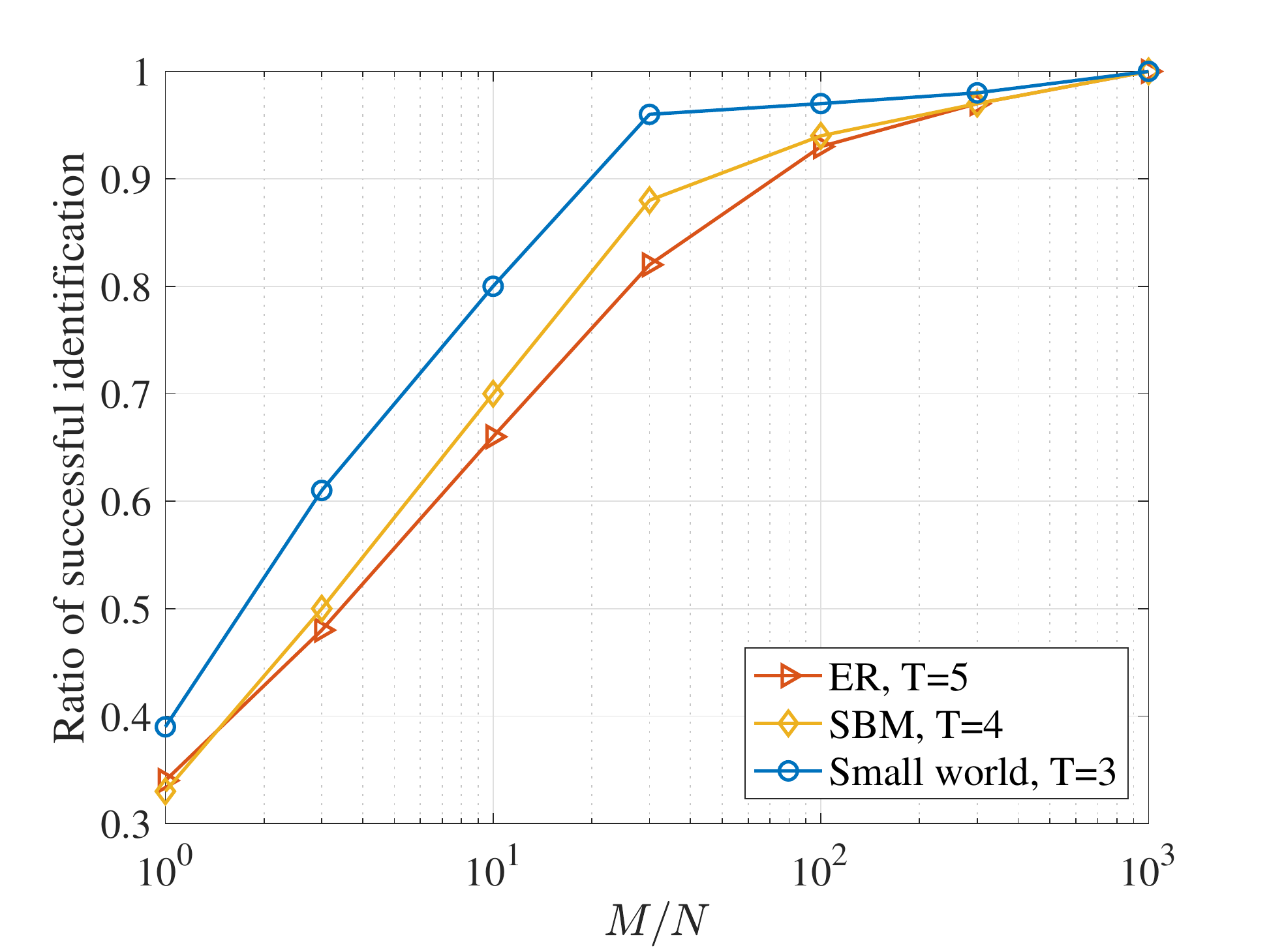}
		\label{Fig:T_est_prob}
	}\\
	\subfigure[]{
		\centering
		\includegraphics[width=0.4\textwidth]{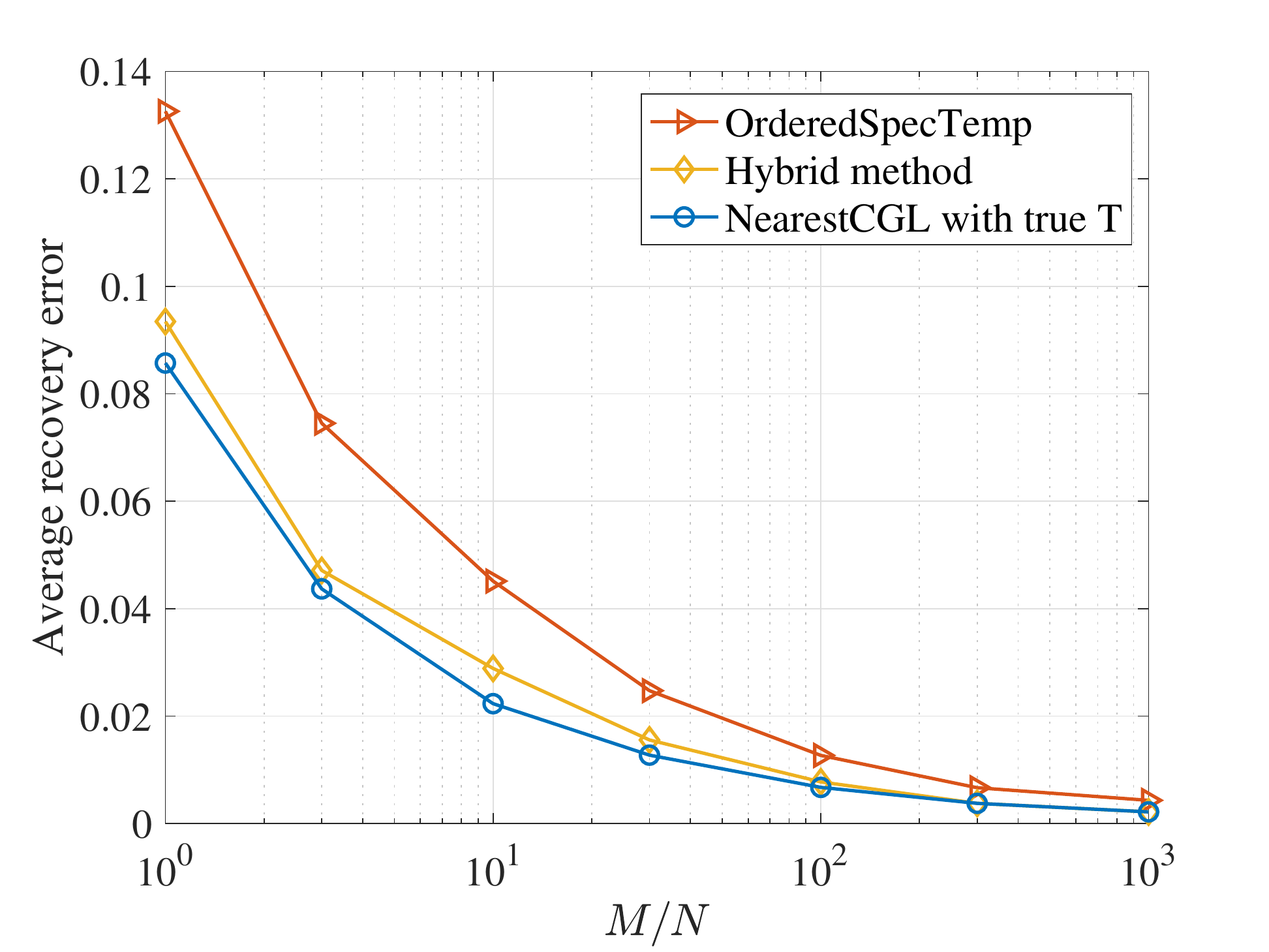}
		\label{Fig:T_est_err}
	}
	\caption{\small (a) Ratio of successful identification of the observation time $T$ for three types of graphs. (b) Average recovery error from a single consensus dynamics with time-invariant diffusion rate and unknown observation time $T$. The proposed hybrid method achieves performance close to the benchmark where $T$ is assumed to be known.}
\end{figure}

In evaluating the performance of the method proposed in Section~\ref{SS:solution_p2}, we first analyze the accuracy of the estimator $\hat{T}$ of the observation time [cf.~\eqref{E:T_est}].
We consider three parameter settings: (i) $T=5$ and ER graphs of size $N=36$ and edge-formation probability $p=0.2$; 
(ii) $T=4$ and SBM graphs with $N=36$ nodes and two blocks where the vertex attachment probabilities across blocks and within blocks are $p_1=0.1$ and $p_2=0.3$, respectively; and 
(iii) $T=3$ and small-world graphs generated via Watts-Strogatz model with $N=36$ nodes, mean node degree $4$ and rewiring probability $0.2$. 
For all settings, {we adopt unweighted graphs and} the diffusion rate is chosen as $\alpha = 0.8/\lambda_{\max}(\mathbf{L})$.
We set $T_{\mathrm{max}} = 10$ in \eqref{E:T_est}.
In solving \eqref{E:T_est}, we must first obtain ${\mathbf{L}}^*_{\mathrm{ord}}$ via OrderedSpecTemp.
Consequently, in \eqref{E:solution_p3} we set $\eta=1$, $d(\mathbf{L},\mathbf{K})=\|\mathbf{L}-\mathbf{K}\|_2$, and $\epsilon$ equal to the smallest possible value in the set $\{0.002\!:\!0.002\!:\!0.03\}\cup\{0.03+0.005\, r \,|\,r=1,2,\cdots\}$ that guarantees feasibility of \eqref{E:solution_p3}. One iteration of a reweighted $\ell_1$ minimization scheme \cite{reweighted} is adopted.
For each setting, we implement $100$ realizations and compute the ratio of successful recovery of $T$.
The results are shown in Fig.~\ref{Fig:T_est_prob}.
We can see a sharp increase on the probability of successful identification with increasing the sample size, eventually converging to $1$. 

Furthermore, for setting (i), we compare the recovery performance of three methods, namely, OrderedSpecTemp (Algorithm~\ref{A:orderedspectemp}), the proposed hybrid algorithm (Algorithm~\ref{A:hybrid}), and NearestCGL (Algorithm~\ref{A:NearestCGL}) with the true $T$.
Notice that this latter method is not applicable in practice (since $T$ is unknown) but rather serves as a benchmark for ideal performance.
{In NearestCGL as well as the hybrid method, we simply set $\beta=0$.}
The results (averaged across $30$ realizations) are shown in Fig.~\ref{Fig:T_est_err}.
We can see that the performance of the proposed hybrid approach is better than that of OrderedSpecTemp, effectively corroborating that $\bbL^*$ is a better estimate than $\bbL^*_{\mathrm{ord}}$ in Algorithm~\ref{A:hybrid}. 
Moreover, the hybrid approach is very close to the benchmark of NearestCGL with perfect knowledge, especially in the large sample size regime.

\subsection{Experiments on Problem~\ref{P:multiple}}\label{Ss:num_exp_p3}

\begin{figure*}
	\centering
	\subfigure[SpecTemp+LEigVec]{
			\centering
			\includegraphics[scale=0.28]{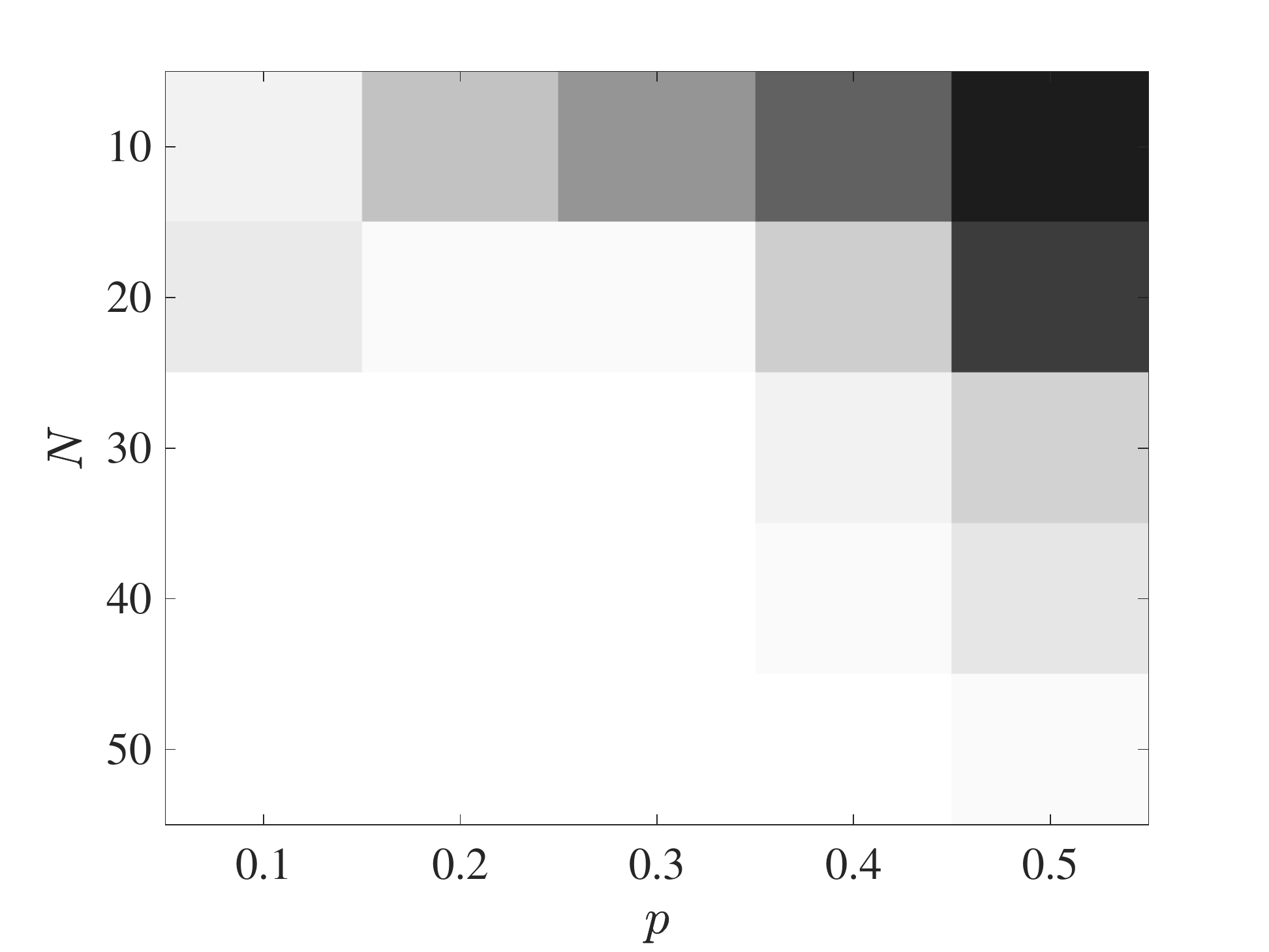}
	}
	\subfigure[OrderedSpecTemp]{
			\centering
			\includegraphics[scale=0.28]{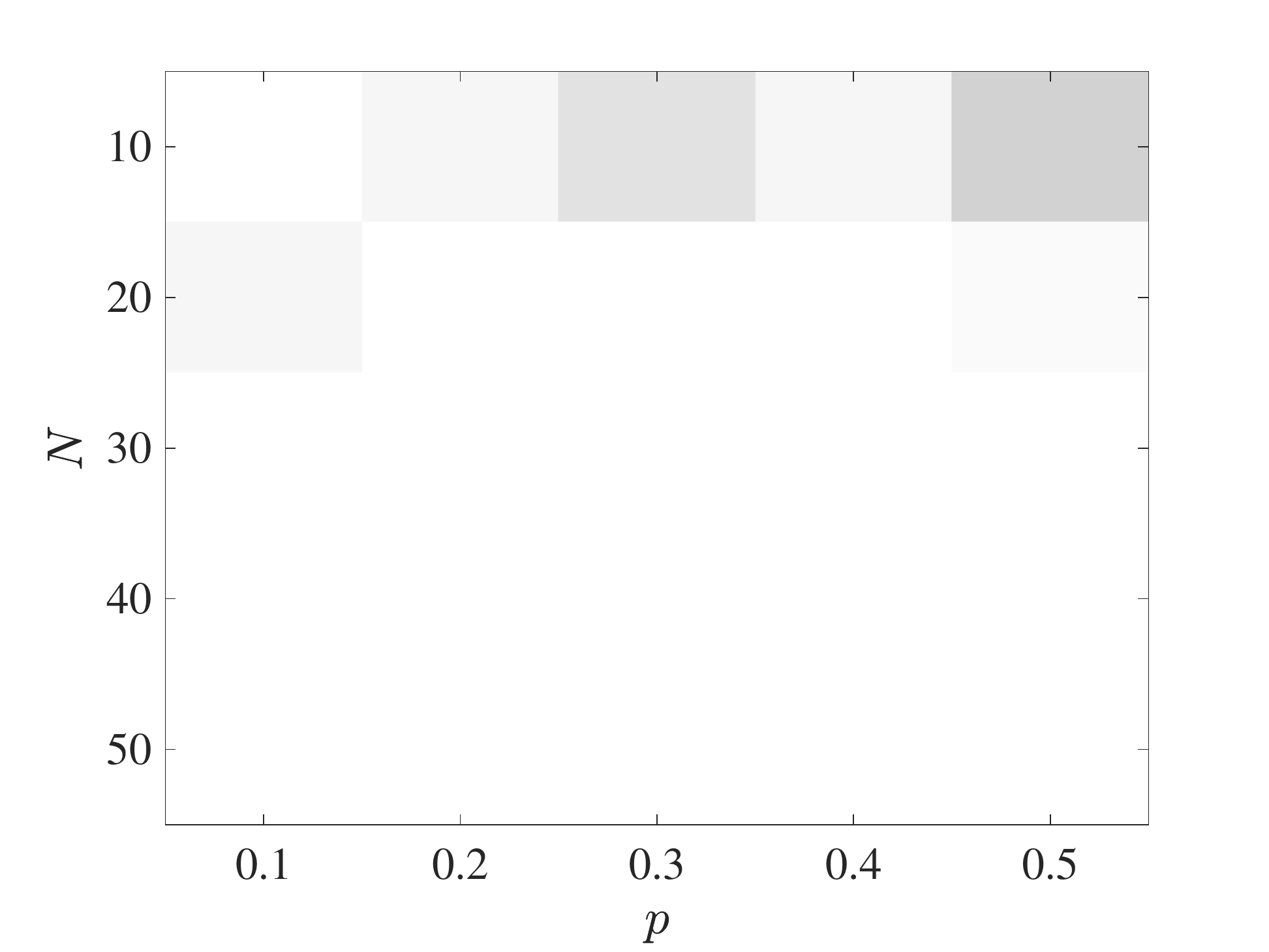}
	}
	\subfigure[The method in~\cite{Segarra2017}]{
			\centering
			\includegraphics[scale=0.28]{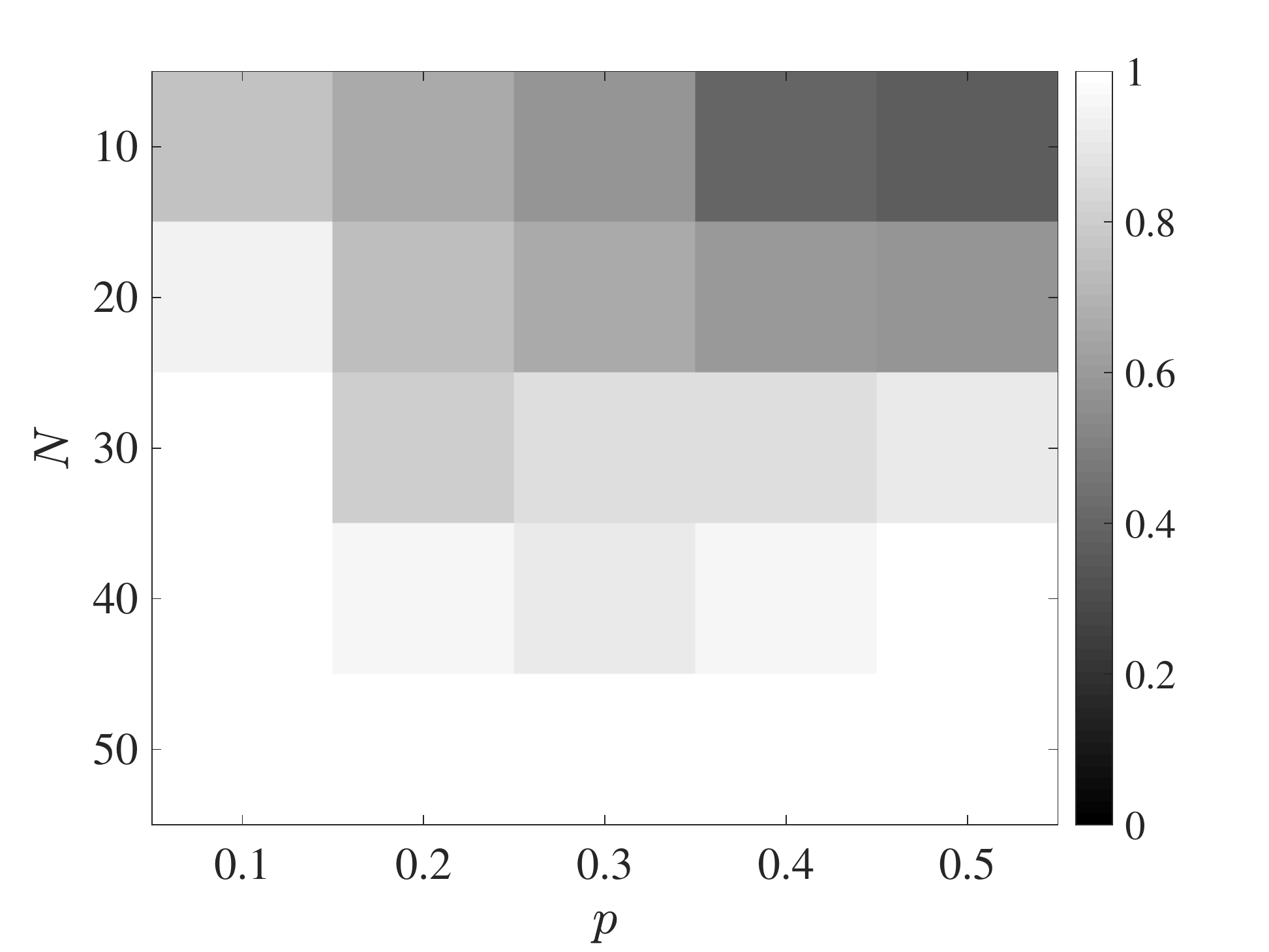}
	}
\vspace{-2mm}
	\caption{\small Comparison of recovery rate between three methods for ER graphs as a function of graph size $N$ and edge-formation probability $p$. The proposed OrderedSpecTemp method outperforms both a variant ignoring the order of the eigenvectors and the method proposed in~\cite{Segarra2017}.}
	\label{Fig:heatmap}
\end{figure*}

\begin{figure}
	\centering
	\subfigure[]{
		\centering
		\includegraphics[scale=0.385]{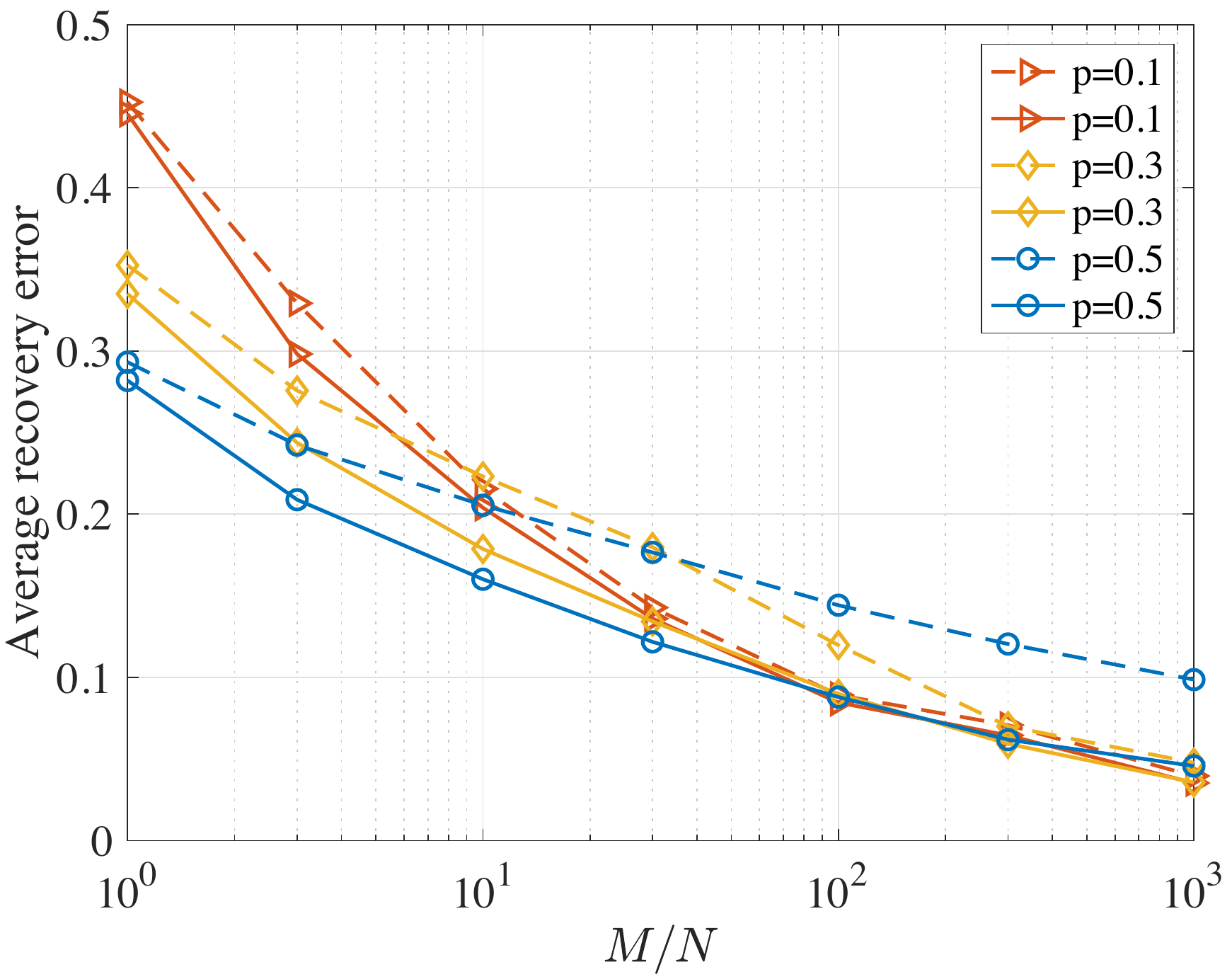}
	}
	
	\subfigure[]{
		
		\centering
		\includegraphics[scale=0.385]{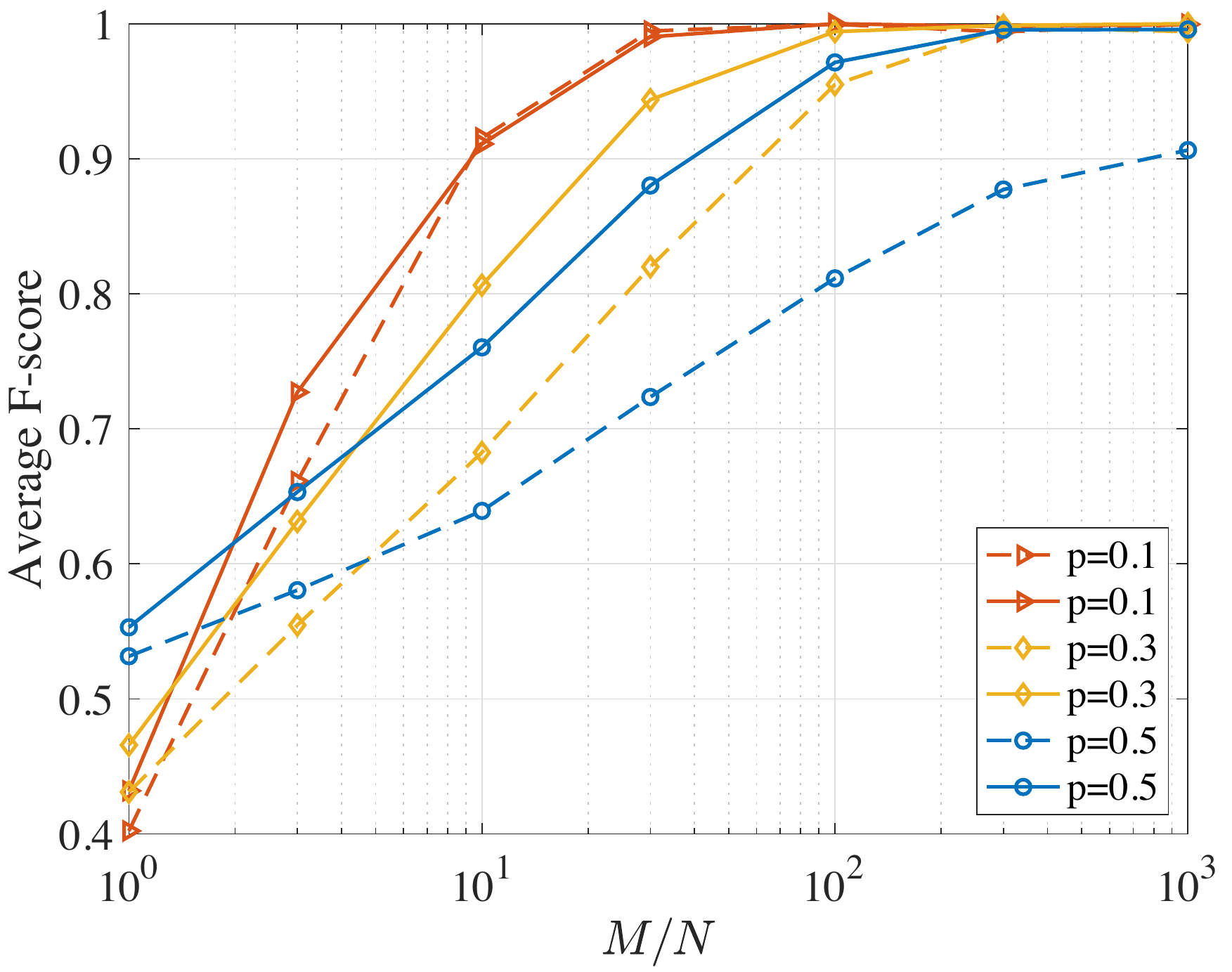}
	}
	\vspace{-2mm}
	\caption{\small Network inference performance under the setting of Problem~\ref{P:multiple} as a function of the number of samples for different graph densities $p$. Solid (dashed) lines correspond to the OrderedSpecTemp (SpecTemp+LEigVec) method. The performance is measured in terms of (a) the average recovery error, and (b) the average F-score.}
	\vspace{-4mm}
	\label{Fig:multi_sample_size}
\end{figure}

We first test the proposed OrderedSpecTemp method (Algorithm~\ref{A:orderedspectemp}) when the eigenbasis $\mathbf{V}$ is perfectly known, i.e., when $\bbU = \bbV$ in \eqref{constraint2}. 
This situation corresponds to the infinite sample size regime (cf. Theorem~\ref{T:eigenvector_converge}).
A high recovery rate under this setting is important since it is a necessary condition for acceptable recovery in the finite sample size regime.
We compare OrderedSpecTemp with two other methods: 
(i) SpecTemp+LEigVec: in this method, we assume that the full order information is unavailable and only the index of the leading eigenvector is known, so we construct the problem similar to \eqref{E:solution_p3} while in \eqref{constraint3} we only keep the constraint $\gamma_N=1$;
(ii) The method proposed in \cite{Segarra2017}, where the constraint \eqref{constraint3} is replaced by $\gamma_i\leq\gamma_{i+\eta}-\epsilon_2 \text{ for } i=1,\cdots,N-\eta$, and $\epsilon_2>0$ can be chosen freely since it only affects the scale of the recovered Laplacian. One advantage of \eqref{constraint3} over the method in~\cite{Segarra2017} is that \eqref{constraint3} can be directly adopted in the case when the CGL $\mathbf{L}$ has repeated eigenvalues while~\cite{Segarra2017} requires the assumption that $\mathbf{L}$ has distinct eigenvalues.

We consider random unweighted ER graphs of varying sizes $N\in\{10,20,\cdots,50\}$ and edge-formation probabilities $p\in\{0.1,0.2,\cdots,0.5\}$.
Since the method in \cite{Segarra2017} is proposed under the assumption that the CGL has distinct eigenvalues, we generate graphs satisfying $\min_{i\neq j}|\lambda_i-\lambda_j|\geq 10^{-4}$.
We set parameters in \eqref{E:solution_p3} as $\mathbf{U}=\mathbf{V}$, $\epsilon=0$ and $\eta=1$.
For the method in \cite{Segarra2017}, we set $\epsilon_2=1$.
{For all three methods, three iterations of a reweighted $\ell_1$ minimization scheme \cite{reweighted} are adopted.}
For each graph generated, we consider the recovery to be successful if $\|{\mathbf{L}}^*-\mathbf{L}\|_{\mathrm{F}}/\|\mathbf{L}\|_{\mathrm{F}}<0.02$.
{Note that the estimate $\mathbf{L}^{\ast}$ here is scaled to have the same trace as the true $\mathbf{L}$ in order to compute the recovery error.} 
The results (averaged across $50$ realizations) are shown in Fig.~\ref{Fig:heatmap}.
We can see that the proposed OrderedSpecTemp method outperforms the other two methods, and its recovery rate is equal to or near $1$ for all graph settings.
Notice that the SpecTemp+LEigVec method works better for sparser graphs. This indicates that the eigenvector order information (ignored by SpecTemp+LEigVec) becomes more important for denser graphs.

We now test OrderedSpecTemp in the finite sample size regime, and compare it with SpecTemp+LEigVec to further study the value of incorporating the eigenvector ordering information.
We consider unweighted ER graphs of size $N=36$ and varying edge-formation probabilities $p=\{0.1,0.3,0.5\}$.
For each graph, our goal is to recover the CGL $\mathbf{L}$ from the observation of $M$ synthetic consensus dynamics, where we vary $M$ from $N$ to $10^3 N$.
For each dynamics, we draw the input from a standard multivariate Gaussian distribution, selecting $T_k$ uniformly at random in $\{3,4,5\}$ and each diffusion rate $\alpha^{(k)}_t$ uniformly at random in $(0,1/\lambda_{\max}(\mathbf{L}))$.
We set $\eta=1$ for OrderedSpecTemp, {use the spectral norm as the matrix distance}, and set $\epsilon$ as the smallest possible value in the set $\{0.005 \, r \,|\,r=1,2,\cdots\}$ that guarantees feasibility of the optimization problem for both two methods [cf.~\eqref{E:solution_p3}].
Apart from the recovery error, we also consider the F-score metric, which is commonly used to evaluate binary classification performance and computed as
\begin{equation*}\label{E:f_score}
	\text{F-score}({\mathbf{L}}^*,\mathbf{L}) = \frac{2 \mathrm{tp}}{2\mathrm{tp}+\mathrm{fn}+\mathrm{fp}},
\end{equation*}
where $\mathrm{tp}$, $\mathrm{fp}$ and $\mathrm{fn}$ respectively represent true-positive, false-positive and false-negative detection of edges in the estimate ${\mathbf{L}}^*$ with respect to edges in the true $\mathbf{L}$.
F-score takes values between $0$ and $1$, and $1$ represents perfect classification.
The results (averaged across $20$ realizations) are shown in Fig.~\ref{Fig:multi_sample_size}.

\begin{figure*}
	\centering
	\subfigure[]{
			\centering
			\includegraphics[width=0.30\textwidth]{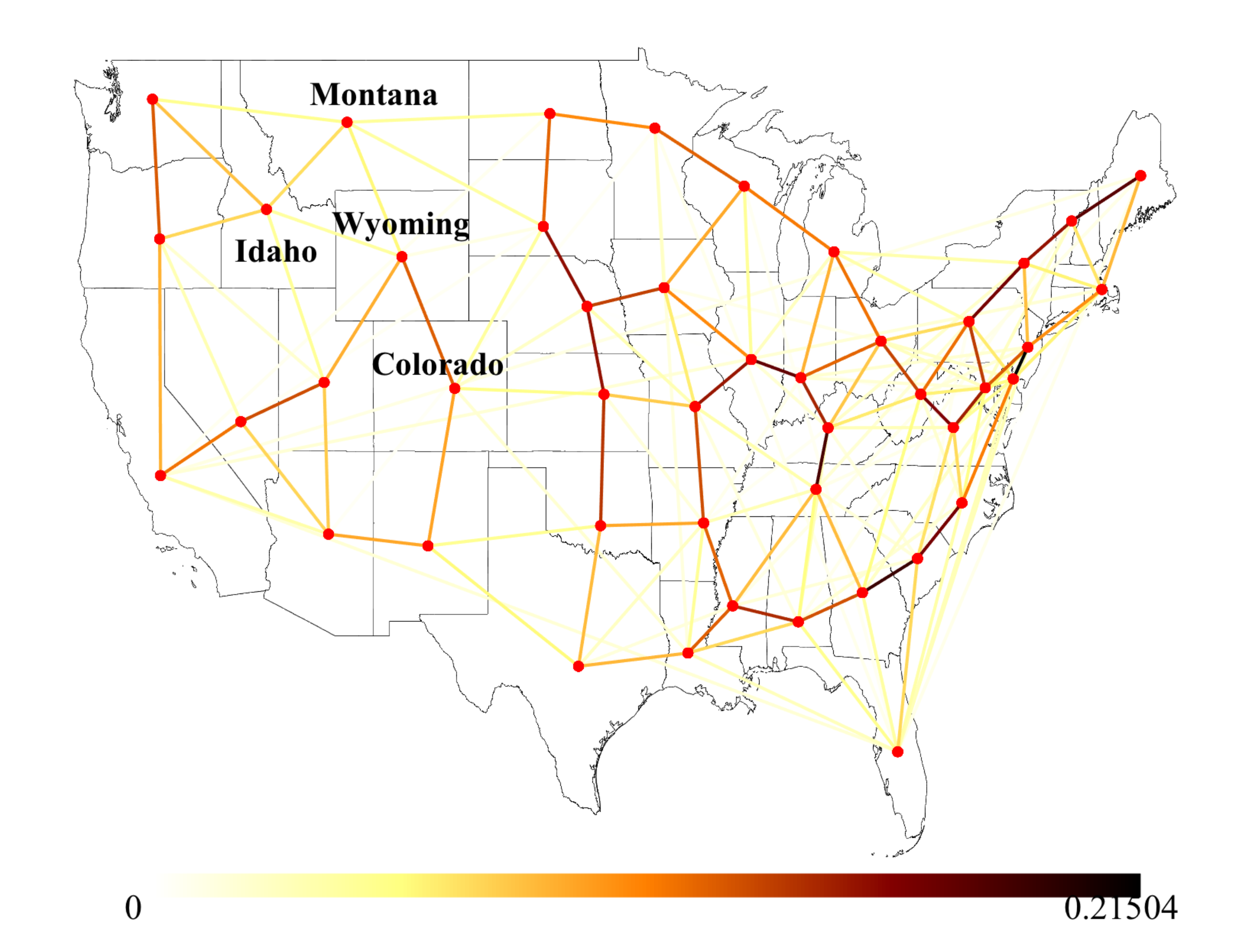}
			\label{Fig:us_temp}
		}	
			\subfigure[]{
			\centering
			\includegraphics[width=0.30\textwidth]{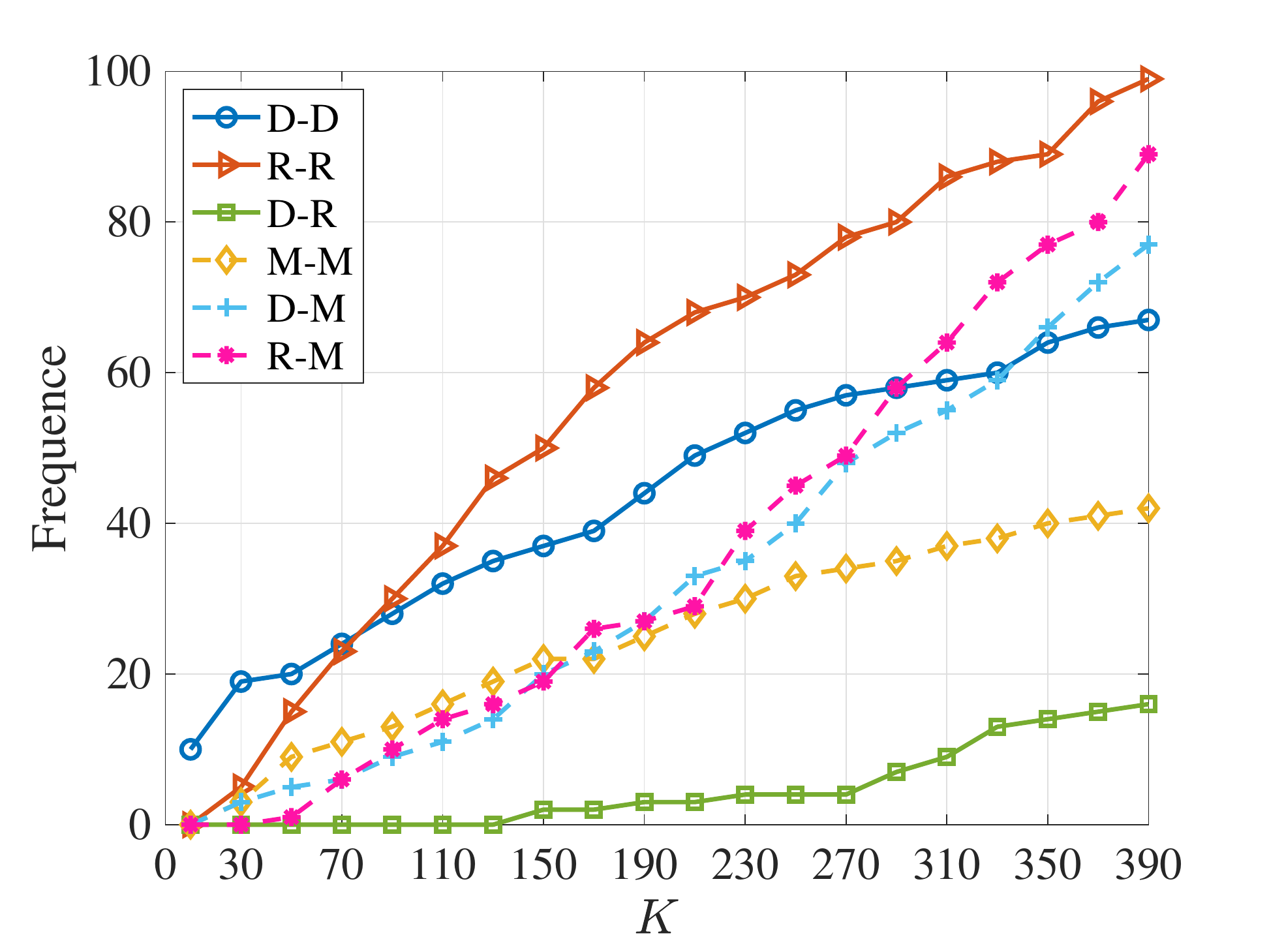}
			\label{Fig:edge_count}
		}
			\subfigure[]{
			\centering
			\includegraphics[width=0.30\textwidth]{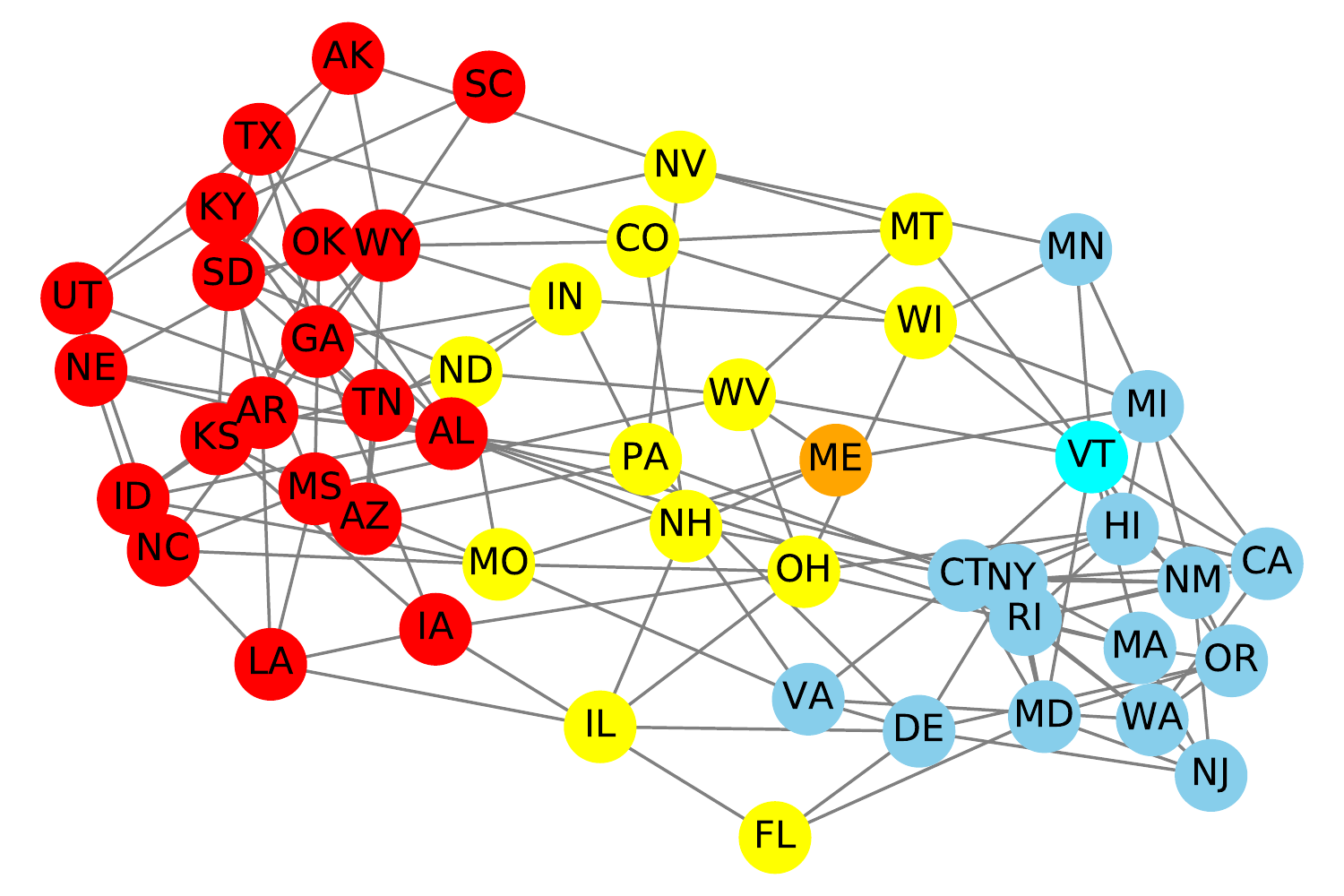}
			\label{Fig:top150}
		}	
	\vspace{-2mm}
	\caption{\small 
	(a) Network topology inference from temperature data using the OrderedSpecTemp method. The edges are colored so that darker colors represent larger edge weights. The east-west split by the Rocky Mountains is captured by the low weights of the recovered edges in the region. 
	(b) Frequency of edges in the senate network that are within and across party classes. As expected, ties within the Democratic and Republican Parties are stronger and more numerous.
	(c) Senate network recovered with top-$150$ edges sorted by weight. Red/blue nodes denote states whose two senators are both republican/democratic; yellow nodes denote states having one republican senator and one democratic senator; the orange/cyan node denotes the state having one republican/democratic senator and one independent senator.}
\vspace{-2mm}
\end{figure*}

As displayed in Fig.~\ref{Fig:multi_sample_size}, for both methods and all graph density settings, we observe a monotonous decrease of the recovery error and a monotonous increase of the F-score with increasing number of samples $M$.
This is not surprising since we know that for larger sample size, the eigenbasis of the sample covariance becomes closer to the eigenbasis of the CGL and thereby facilitates recovery.
We can also see that, OrderedSpecTemp outperforms SpecTemp+LEigVec and the performance gap increases when the graph becomes dense.
This is consistent with the results in Fig.~\ref{Fig:heatmap}, i.e. SpecTemp+LEigVec works well for very sparse graphs, while when the graph density increases, the full order information becomes more valuable in aiding recovery.

\subsection{Topology inference from real-world data}\label{Ss:num_exp_real}

We present two different real-world case studies where we recover connections between the states of the U.S. in both cases but based on very different information sources, namely, temperature measurements and congressional roll-call votes.

\vspace{1mm} \noindent {\bf Temperature network.}
We apply the proposed OrderedSpecTemp method on a real-world dataset consisting of $M=5844$ average daily temperature measurements collected from $N=45$ states in the U.S. over the years $2000$-$2015$ \cite{us_temp_data}.
This same dataset was analyzed in~\cite{Hilmi2018}. 
The temperature signals are spatially smooth across different states, and the Rocky Mountains region (which is mainly located in the states of Montana, Wyoming, Colorado, and Idaho) has lower average temperature values than its geographical neighborhood (cf. Fig.~7 in~\cite{Hilmi2018}).

In solving \eqref{E:solution_p3}, we set $\eta=1$, {$d(\mathbf{L},\mathbf{K})=\|\mathbf{L}-\mathbf{K}\|_{2}$,} and $\epsilon$ equal to the smallest possible value (found via five iterations of binary search between $0$ and $1$) that guarantees feasibility of \eqref{E:solution_p3}. The result shown in Fig.~\ref{Fig:us_temp} is obtained after three iterations of {a reweighted $\ell_1$ minimization scheme }~\cite{reweighted}.

As shown in Fig.~\ref{Fig:us_temp}, the edges connecting neighboring states generally have larger weights since temperature values are similar between regions located near to each other.
In this sense, temperature values are accurately described by a consensus dynamics where discrepancies between neighbors tend to be reduced.
However, there are other factors -- besides the geographical distance -- that can also influence the similarity of temperature values, such as landform and altitude.  
It can be observed that the edges between the Rocky Mountain states and their neighbors to the east have relatively small weights.
In this way, the recovered network captures the natural barrier for temperature similarity imposed by the mountain formation.

As mentioned, in \cite{Hilmi2018} the authors apply their proposed method -- referred to as StructGLasso in this paper -- to this same dataset by assuming different graph filter functions and parameters.
The network recovered here is comparable to the one inferred in \cite{Hilmi2018} for an exponential decay filter, which was deemed as better revealing the structure of the signal (cf. Fig. 8 (d)-(f) in \cite{Hilmi2018}).
This implies that, in the absence of prior knowledge about the specific filter $h(\mathbf{L})$ that is driving the underlying network process, adopting a more versatile model where only a decaying frequency response is assumed [cf.~\eqref{constraint3}] can be beneficial.
On the contrary, if the filter type is known in advance, then it is reasonable to incorporate it in the recovery method, as advocated in~\cite{Hilmi2018}.

\vspace{1mm} \noindent {\bf Senate network.}
We now apply OrderedSpecTemp on a real-world dataset from congressional roll-call votes in the U.S.~\cite{voting_data}.
We use the roll-call data of $100$ senators ($2$ per state) in the $114$th congress ($2015$-$2017$), consisting of $M=502$ roll-calls.
We quantify \emph{yea}, \emph{nay} and other cases (e.g. abstention) as $1$, $-1$ and $0$, respectively, to represent each senator's opinion.
Since each state has two senators, we use the sum of their opinions as the graph signal value of their state, resulting in a graph with $N=50$ nodes.
We choose the parameters in \eqref{E:solution_p3} following the same process explained for the temperature network.

We divide the states into three categories: (i) labeled as D if both senators in this state are from the Democratic Party, (ii) labeled as R if both senators in this state are from the Republican Party, (iii) labeled as M if the senators are from different (mixed) parties. 
The recovered network contains $391$ edges, and we count the number of edges inside each category and between different categories in the top-$K$ edges sorted by weight for different $K$, as shown in Fig.~\ref{Fig:edge_count}.
It can be observed that there are more edges inside the categories D and R, while less edges between them.
In Fig.~\ref{Fig:top150}, we plot the recovered network with the top-$150$ edges in spring layout using NetworkX~\cite{networkx}.
It can be observed that two tight clusters of states emerge (D and R) with a looser cluster of mixed states M connecting these two.

\begin{figure}
	\centering
	\centering
	\includegraphics[scale=0.385]{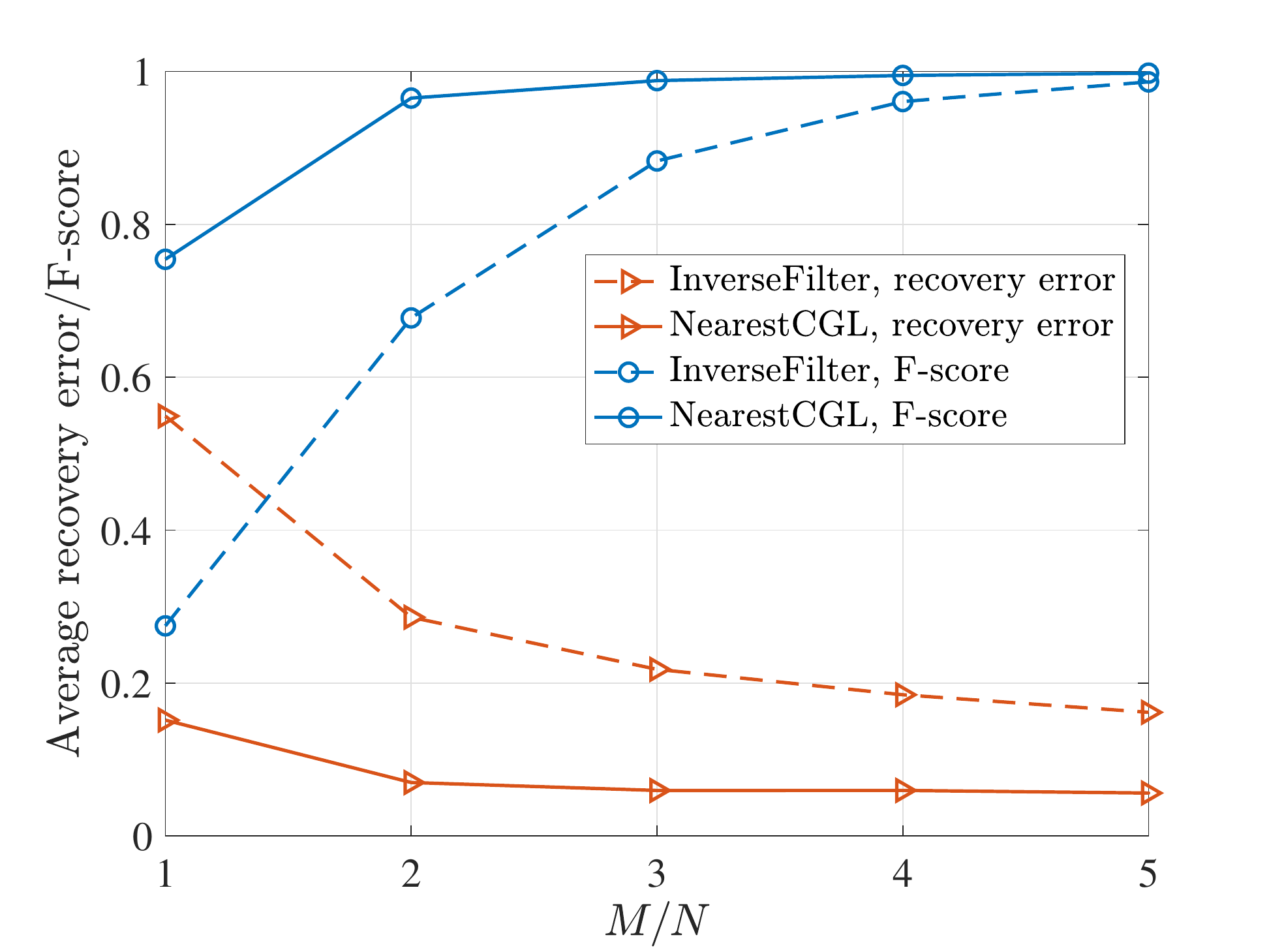}
	\vspace{-2mm}
	\caption{\small {Average recovery error and F-score as a function of the number of samples under the setting of Problem~\ref{P:single} for the U.S. political blogs network of $1,\!222$ nodes.}}
	\vspace{-4mm}
	\label{Fig:large-graph}
\end{figure}

\subsection{Scalability of the proposed algorithms}\label{SS:scaling}

To illustrate the scalability of the proposed algorithms, we infer a large-scale real-world network under the setting of Problem~\ref{P:single} [cf.~\eqref{E:fast2}].
In particular, we consider the U.S. political blogs network\footnote{The data can be found in \url{http://www-personal.umich.edu/~mejn/netdata/}.} which is a directed network of hyperlinks between weblogs on US politics~\cite{uspblog}.
We preprocessed the graph by: (i) converting the directed graph to an undirected one, (ii) removing self-loops and multi-edges, and (iii) keeping the largest connected component.
After these changes, we obtain an undirected and unweighted graph with $1,\!222$ nodes and $16,\!714$ edges.
We adopt a graph filter of the form $h(\mathbf{L})=(\mathbf{I}-\alpha\mathbf{L})^{T}$, where we set $\alpha=0.9/\lambda_{\max}(\mathbf{L})$ and $T=15$.
We consider different values for the ratio of sample size and graph size, i.e., $M/N=1,2,3,4,5$.
The corresponding values for the regularization parameter are selected as $\beta=40,30,25,20,5$, respectively.
The results, averaged across 20 realizations, are shown in Fig.~\ref{Fig:large-graph}.
We observe that the proposed NearestCGL outperforms the InverseFilter algorithm both in terms of recovery error and F-score.
As an indication of scalability, we remark that solving the problem~\eqref{E:fast2} once for this specific graph requires around $17$ seconds of computation in a standard laptop computer. 


\section{Conclusions}\label{S:conclusions}


We proposed a set of algorithms for the identification of a network based on observing snapshots of consensus processes considering different levels of parameter uncertainty. 
To achieve this, we constructed convex optimization problems that output a sparse, valid graph Laplacian which is provably consistent with the spectral information obtained from the observed data.
Finally, we showcased the effectiveness of the proposed methods in synthetic and real-world scenarios.

 
Potential future research avenues include: (i) investigation of different types of restrictions on the available data, such as observation models with snapshot data sampled from partial nodes of the graph; (ii) study of the trade-off between specific network topologies and the required sample size to achieve a desired level of estimation accuracy; (iii) consideration of a richer class of dynamical models, including non-deterministic processes such as switched systems; (iv) joint estimation of several related networks from the concurrent observation of consensus dynamics~\cite{segarra_2017_joint, zhu_2019_estimation}; {and (v) topology inference of directed graphs.}


\begin{appendices}

\section{Random variables and properties}\label{A:random_variable}

In this appendix, we review concepts and properties of random variables that we used throughout the paper; for a thorough review, see~\cite{random_variable}.

\begin{mydefinition}[Sub-Gaussian random variable]\label{D:subgaussian}
A random variable $x$ with mean $\mu=\mathbb{E}[x]$ is sub-Gaussian if there is a positive parameter $\sigma$ such that 
\begin{equation}
\mathbb{E}[e^{\gamma(x-\mu)}]\leq e^{\frac{\sigma^2\gamma^2}{2}} \quad \text{for all } \gamma\in\mathbb{R}.
\end{equation}
\end{mydefinition}

\begin{mylemma}[Sub-Gaussian tail bound~\cite{random_variable}]\label{L:subgaussian}
Suppose that $x$ is sub-Gaussian with mean $\mu$ and parameter $\sigma$, then
\begin{equation}
\mathbb{P}[|x-\mu|\geq l] \leq 2e^{-\frac{l^2}{2\sigma^2}} \quad\text{for all } l\in\mathbb{R}.
\end{equation}
\end{mylemma}

\begin{mydefinition}[Sub-exponential random variable]
A random variable $x$ with mean $\mu=\mathbb{E}[x]$ is sub-exponential if there are positive parameters $(\nu,b)$ such that 
\begin{equation}
\mathbb{E}[e^{\gamma(x-\mu)}]\leq e^{\frac{\nu^2\gamma^2}{2}}\quad \text{for all } |\gamma|<1/b. 
\end{equation}
\end{mydefinition}

\begin{mylemma}[Sub-exponential tail bound~\cite{random_variable}]\label{subexp_tail}
Suppose that $x$ is sub-exponential with mean $\mu$ and parameters $(\nu,b)$, then 
\begin{equation}
\mathbb{P}[|x-\mu|\geq l]\leq\left\{
\begin{array}{ll}
2e^{-\frac{l^2}{2\nu^2}}  & \text{if }\, 0\leq l\leq \frac{\nu^2}{b},\\
2e^{-\frac{l}{2b}}     &\text{if }\,  l>\frac{\nu^2}{b}.
\end{array} \right. 
\end{equation}
\end{mylemma}

\begin{mylemma}[\!\!\cite{Segarra2017}]\label{product_gaussian}
Let $x\sim\mathrm{N}(0,\sigma_1^2)$ and $y\sim\mathrm{N}(0,\sigma_2^2)$ be independent random variables. 
Then their product $z=xy$ is sub-exponential with parameters $(\sqrt{2}\sigma_1\sigma_2,\sqrt{2}\sigma_1\sigma_2)$.
\end{mylemma}

\begin{mylemma}[\!\!\cite{Segarra2017}]\label{square_gaussian}
Let $x\sim\mathrm{N}(0,\sigma^2)$, then its square $z=x^2$ is sub-exponential with mean $\mathbb{E}(z)=\sigma^2$ and parameters $(2\sigma^2,4\sigma^2)$.
\end{mylemma}

\begin{mylemma}[\!\!\cite{random_variable}]\label{sum_subexp}
Let $\{x_k\}_{k=1}^M$ be independent sub-exponential random variables with mean $\mu_k$ and parameters $(\nu_k,b_k)$ respectively. 
Then the variable $\sum_{k=1}^M (x_k-\mu_k)$ is sub-exponential with parameters $(\sqrt{\sum_{k=1}^M \nu_k^2}, \,\max\limits_{k} b_k)$.
\end{mylemma}

\end{appendices}

\bibliographystyle{IEEEtran}
\bibliography{GL_consensus}

\end{document}